\documentclass[12pt,USenglish]{article}
\pdfoutput=1

\usepackage{setspace}
\usepackage{cite}

\renewcommand{\Phi}{\phi}

\usepackage[utf8]{inputenc}
\usepackage{geometry}
\usepackage[USenglish]{babel}
\usepackage{verbatim}
\usepackage{prettyref}
\usepackage{amsmath}
\usepackage[normalem]{ulem}
\usepackage{amssymb}
\usepackage{graphicx}
\usepackage{setspace}
\usepackage{esint}

\usepackage{color}
\definecolor{darkgreen}{rgb}{0,0.5,0}
\definecolor{darkblue}{rgb}{0,0,0.6}
\definecolor{purple}{rgb}{0.4,.2,0.7}
\definecolor{orange}{rgb}{0.95, 0.5, 0.3}

\usepackage[colorlinks=true,citecolor=darkgreen,linkcolor=darkblue,urlcolor=blue]{hyperref}

\numberwithin{equation}{section}
\numberwithin{figure}{section}
\numberwithin{table}{section}

\def\red#1{{\color{red} #1}}

\def\be{\begin{equation}}
\def\ee{\end{equation}}
\def\bea{\begin{eqnarray}}
\def\eea{\end{eqnarray}}
\def\ba{\begin{align}}
\def\ea{\end{align}}
\def\d{\text{d}}

\def\z{\zeta}

\def\d{\text{d}}
\def\id{1_{2\times 2}}

\newcommand{\bz}{\bar{z}}

\newcommand{\V}{\mathcal{V}}

\begin{document}
\begin{spacing}{1.3}
\begin{flushright}
{\tt MIT-CTP-4780}
\end{flushright}

~
\vskip5mm

\begin{center} {\Large \bf Black Hole Collapse in the $1/c$ Expansion}

\vskip10mm

Tarek Anous,$^{1}$ Thomas Hartman,$^{2}$ Antonin Rovai$^{1,3}$ \& Julian Sonner$^{1,3}$\\
\vskip1em
{\it 1) Center for Theoretical Physics, Massachusetts Institute of Technology, Cambridge, MA 02139, USA} \\
\vskip5mm
{\it 2) Department of Physics, Cornell University, Ithaca, New York, USA} \\
\vskip5mm
{\it 3) Department of Theoretical Physics, University of Geneva, 25 quai Ernest-Ansermet, 1214 Gen\`eve 4, Switzerland} 
\vskip5mm

\tt{ tanous@mit.edu, hartman@cornell.edu, \{antonin.rovai,julian.sonner\}@unige.ch}

\end{center}

\vskip10mm

\begin{abstract}
We present a first-principles CFT calculation corresponding to the spherical collapse of a shell of matter in three dimensional quantum gravity.
In field theory terms, we describe the equilibration process, from early times to thermalization, of a CFT following a sudden injection of energy at time $t=0$.
By formulating a continuum version of Zamolodchikov's monodromy method to calculate conformal blocks at large central charge $c$, we give a framework to compute a general class of probe observables in the collapse state, incorporating the full backreaction of matter fields on the dual geometry. This is illustrated by calculating a scalar field two-point function at time-like separation and the time-dependent entanglement entropy of an interval, both showing thermalization at late times. The results are in perfect agreement with previous gravity calculations in the AdS$_3$-Vaidya geometry. Information loss appears in the CFT as an explicit violation of unitarity in the $1/c$ expansion, restored by nonperturbative corrections.  
\end{abstract}


\pagebreak 
\pagestyle{plain}

\setcounter{tocdepth}{2}
{}
\vfill
\tableofcontents
\newpage
\section{Introduction}

The contradiction between black holes and local quantum field theory, manifested in the information paradox and related puzzles, is sharpest for transient black holes that form by collapse, slowly evaporate, and eventually disappear. In three or more spacetime dimensions, AdS/CFT strongly suggests that information is recovered \cite{Strominger:1996sh,Maldacena:1997re}. In the three-dimensional case, enhanced symmetries greatly simplify the problem of calculating quantum gravity observables, so the 3d BTZ black hole \cite{Banados:1992wn} is perhaps the ideal arena to address information loss. There is every reason to believe that the mechanism for information recovery in 3d gravity is the same as in four dimensions (unlike the 2d case, which is exactly solvable but qualitatively different \cite{Callan:1992rs}); after all, 3d gravity coupled to matter can capture a full higher dimensional string theory \cite{Strominger:1996sh,Maldacena:1997re}. 

Black holes in AdS$_3$ can, in principle, be treated nonperturbatively using the dual CFT. An important first step  is to derive the leading order, semiclassical gravity predictions directly from CFT. Information loss is then a question of nonperturbative corrections to this leading term.

Many of the predictions of pure 3d gravity --- meaning the gravitational sector alone, ignoring the contributions of matter fields --- can already be derived from CFT. Early successes include the calculation of black hole entropy \cite{Strominger:1996sh, Strominger:1997eq}, thermodynamics \cite{Maldacena:1998bw, deBoer:2006vg}, and much more. Recently, these methods have been recast and extended to a set of general techniques for computing observables in large-$c$ conformal field theory, without reference to a particular Lagrangian or other microscopic details \cite{Hartman:2013mia,Barrella:2013wja,Hartman:2014oaa,Fitzpatrick:2014vua}. (See also \cite{Gaiotto:2007xh,Yin:2007gv,Headrick:2010zt,Keller:2011xi} for earlier work in this direction, and \cite{Chen:2013kpa,Chen:2014unl,Chen:2014ehg, Chen:2015kua, Headrick:2015gba,Fitzpatrick:2015zha,Fitzpatrick:2015qma,Fitzpatrick:2015foa,Fitzpatrick:2015dlt, Fitzpatrick:2016thx, Perlmutter:2013paa, Perlmutter:2015iya,Hijano:2015qja,Perlmutter:2016pkf,Asplund:2014coa,Asplund:2015eha,Caputa:2014vaa,Keller:2014xba,Jackson:2014nla,Benjamin:2015hsa,Benjamin:2015vkc,Beccaria:2015shq,Chang:2015qfa,Maldacena:2015iua} for related developments and applications.) This `$1/c$ expansion' reflects the perturbative expansion in $\ell_{\rm Planck}/\ell_{\rm AdS} \sim 1/c$ on the gravity side. It relies on a large central charge $c$ and a sparse spectrum of low-dimension operators, two ingredients universal to every theory with a gravitational dual (as discussed for example in \cite{Callan:1996dv,Heemskerk:2009pn}). In many cases, it also relies essentially on the Virasoro algebra, which is connected to the topological nature of pure 3d gravity.  However, the difficult and interesting questions in quantum gravity, including the information paradox, require coupling gravity to dynamical matter fields so that the theory is no longer topological.  These additional degrees of freedom must ultimately be incorporated into the $1/c$ expansion.

A technique for computing correlation functions of arbitrary heavy operators in the $1/c$ expansion was formulated in \cite{Hartman:2013mia}, using a monodromy prescription that was introduced in classic work of Zamolodchikov \cite{zamolodchikov1986two,zamo}.  `Heavy' means the scaling dimension is $\Delta \gg 1$, including states with $\Delta \sim c$ that backreact on the geometry on the gravity side. The first steps towards coupling gravity to matter, in CFT language, were made in \cite{Fitzpatrick:2014vua,Fitzpatrick:2015zha,Fitzpatrick:2015foa}, where the monodromy method was used to calculate universal long-distance correlators in high-energy eigenstates. The calculations give thermal CFT answers, which agree with the corresponding calculations done in eternal black hole geometries on the gravity side, so these heavy eigenstates are interpreted as black hole microstates.
 Similar methods were used to calculate geodesic lengths and entanglement entropies in eigenstates and local quench states \cite{Asplund:2014coa}. All of these calculations involve a small number of local operator insertions, interpreted on the gravity side as defects propagating on a fixed geometry.

These methods have not yet been applied to collapsing black holes, the most interesting arena for information puzzles. In fact, to our knowledge, there has never been a CFT calculation of dynamical quantities dual to a collapsing black hole in any dimension.  The aim of this paper is to fill this gap.  
We do so by incorporating the simplest form of smooth matter into the $1/c$ expansion: null dust.  Null dust can be created by inserting local operators in the CFT.  By taking the limit of an infinite number of dust particles, holding fixed the total energy, we construct CFT states dual to collapsing black holes.  The limiting procedure replaces the large number of discrete particles by a smooth matter stress tensor supported on a spherically symmetric collapsing null shockwave. It is dual, therefore, to the Vaidya geometry in AdS$_3$. 
This geometry is ideally suited to $1/c$ techniques, since it allows for a study of black hole collapse but is insensitive to the detailed dynamics of the underlying matter fields.

In this dynamical CFT state, we develop large-$c$ methods to compute probe observables, including $n$-point correlation functions and entanglement entropies. Unlike all of the previous $1/c$ calculations described above, the stress tensor in the collapse state is not meromorphic, so this requires essentially new techniques.  The results match precisely with numerous gravity calculations in the literature \cite{Ryu:2006bv,Hubeny:2007xt,AbajoArrastia:2010yt,Balasubramanian:2010ce,Balasubramanian:2011ur,Balasubramanian:2012tu,Ziogas:2015aja}. We also use our CFT methods to predict new observables in AdS$_3$-Vaidya, such as the equal space two-point function at time-like separation in the global geometry. The non-trivial agreement, where gravity answers are known, lends support to the claim that this state is dual to the collapse geometry.  Interestingly, these observables `see'  the geometry behind both the event horizon and the apparent horizon of the collapsing black hole \cite{Hubeny:2007xt,AbajoArrastia:2010yt}.  Such probes have been discussed in CFT before \cite{Fidkowski:2003nf,Hartman:2013qma}, though not in detail for black holes formed by gravitational collapse.

Our primary tool is the Virasoro vacuum block at large $c$.  This fascinating object is, roughly speaking, the sector of the CFT dual to the gravitational sector in the bulk \cite{Hartman:2013mia}. On the one hand, Virasoro blocks are completely fixed by symmetry, but on the other hand we use the vacuum block to extract truly dynamical quantities which are \textit{not} fixed by symmetry. This is possible in theories with a large gap in the spectrum of operator dimensions, by making some reasonable assumptions about the dominant contributions to the correlator in an operator product expansion. In the context of our collapse state this means that we are able to study the nonlinear dynamics of a large number of `constituents'. Such dynamics are clearly not determined by symmetry, although our large-$c$ conformal block techniques form a crucial ingredient. From a more fundamental perspective we thus derive dynamics within a universal sector of 3d quantum gravity with matter which non-trivially matches with semiclassical expectations from Einstein gravity. It is evident from our results that the corresponding correlators in the theory at small $c$ look nothing like semiclassical gravity, even though this case is constrained by conformal symmetry in exactly the same way.

To treat a smooth matter distribution, as in shell collapse, the main technical challenge is to generalize the notion of the Virasoro vacuum block, and the techniques for calculating it, to an infinite number of operator insertions. We show that this problem simplifies dramatically in the final limit, and leads to an intuitive calculation of the block that in many ways resembles the dual gravity calculation.

Of course, reproducing gravity from CFT does not directly address the information paradox.  In fact, the situation is quite the opposite: our CFT calculation \textit{loses} information! In particular, the probe two-point function $G(t_1, t_2)$ computed in the $1/c$ expansion na\"ively decays exponentially at late times, in agreement with the gravity side, but in violation of unitarity. Yet the CFT is in a pure state and the exact evolution is manifestly unitary. This `paradox' is easily traced to the approximation involved in the $1/c$ expansion, since at late times, operator exchanges that were initially exponentially subleading $\sim e^{-S}$ (where $S$ is the entropy) can come to dominate the correlator. This is similar to Maldacena's information puzzle for eternal black holes \cite{Maldacena:2001kr}. It would be interesting to translate Hawking's paradox or the firewall paradox \cite{Almheiri:2012rt} into 2d CFT along similar lines, but these require evaporating black holes at very late times, so go beyond the present paper. Further remarks on information loss and what we may hope to learn from posing these paradoxes in CFT are in the discussion section.

Aside from applications to black holes, our method also provides a new way to study thermalization in quantum field theory.  There are very few situations where thermalization can be studied analytically, especially at strong coupling. A famous exception is the work of Calabrese and Cardy on sudden quenches \cite{Calabrese:2006rx,Calabrese:2016xau}, where the Hamiltonian $H_0$ of a gapped system is suddenly tuned to criticality $H_0 \to H_{\rm CFT}$ at time $t=0$.  This process is modeled by a boundary state \cite{Calabrese:2006rx,Calabrese:2016xau}, which is a state in the CFT with no long-distance correlations at $t=0$.  Our calculation, on the other hand, corresponds to a different type of equilibration, where we start in the CFT vacuum, then at $t=0$ inject a large amount of energy into the system.  The injected matter has only short distance correlations, but unlike a boundary state, the initial state also has the long range correlations that were already present in the vacuum. Thermalization occurs as the injected matter equilibrates. Our calculations produce the detailed correlators throughout this process, from energy injection to complete thermalization. The Cardy-Calabrese calculations were in rational CFT, where individual modes can appear thermal but true thermalization does not occur. Our setup is a strongly coupled non-rational theory with $c>1$, and such 2d CFTs truly thermalize, much like higher dimensional quantum field theories (see \cite{Asplund:2015eha} for a discussion in the context of entanglement). We give explicit formulae for various two-point functions during the collapse, but our methods also allow for the calculation of higher-point functions. It would be interesting, therefore, to apply them to the study of quantum chaos along the lines of  \cite{Roberts:2014ifa,Maldacena:2015waa,Polchinski:2015cea}, but far from equilibrium.

\section{The Collapse State}
\begin{figure}[h!]
\begin{center}
\includegraphics[width=0.4\textwidth]{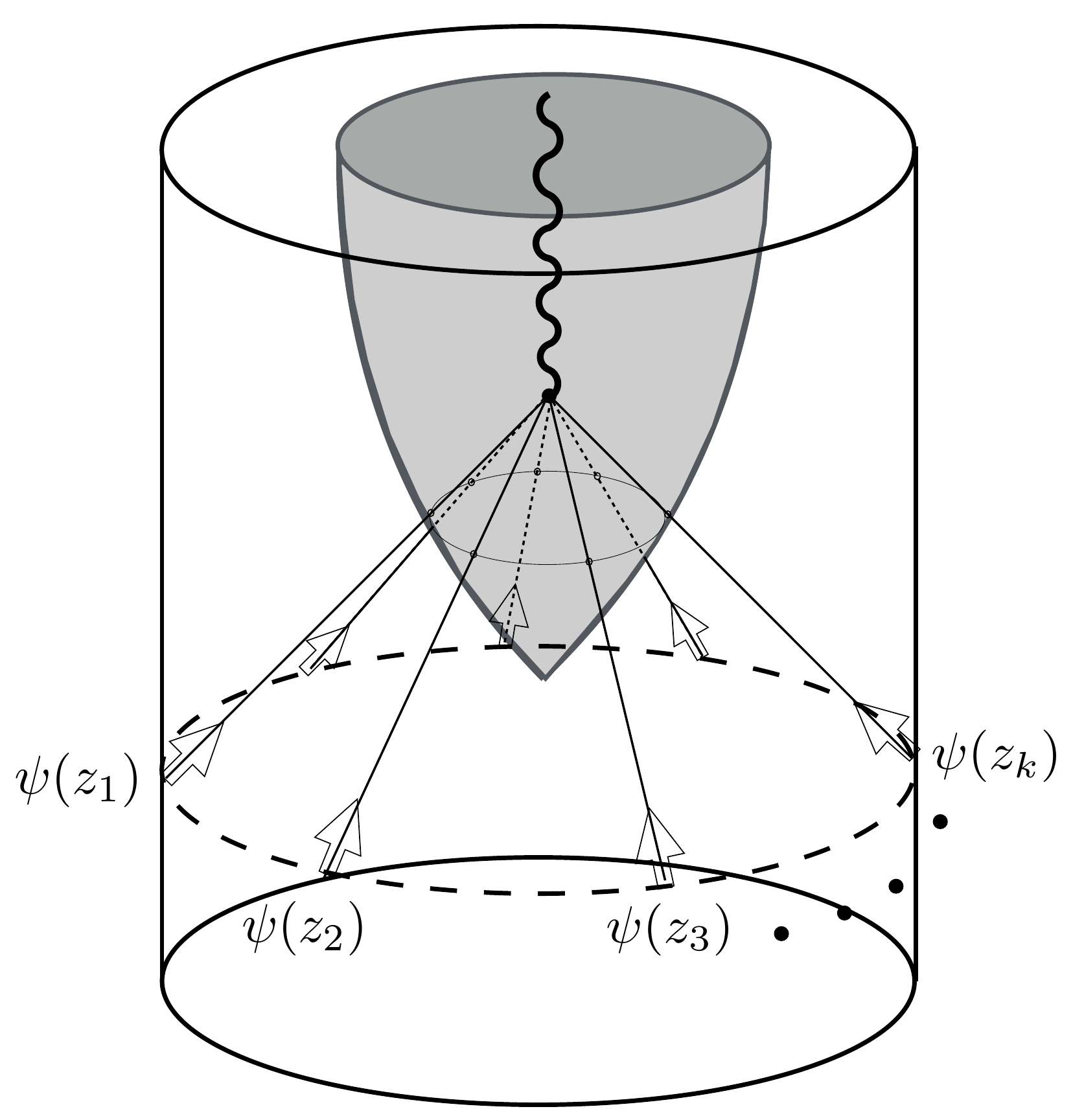}
\caption{\small A shell made up of individual null dust particles collapses to form a BTZ black hole. We have labelled the particles by their dual operator insertion on the boundary in anticipation of our CFT construction in this paper. \label{fig:shellCollapse}}
\end{center}
\end{figure}

\subsection{Motivation from the gravity side}

We will construct a collapsing shell state $|\V\rangle$ in CFT by inserting a large number of local primary operators.  To motivate this construction, we begin by reviewing the well known gravity calculation. The simplest model for black hole formation is the null collapse of a shell of pressureless dust,\footnote{Although we do not consider this possibility here, the 3d black hole can also be formed by colliding a small number of heavy particles \cite{Matschull:1998rv}.  The resulting geometry could be studied in the $1/c$ expansion of the dual CFT, but it lacks spherical symmetry, so we do not expect a simple analytic formula for the probe correlators. The relationship between the colliding particle geometries and spherically symmetric Vaidya collapse was studied recently in \cite{Lindgren:2015fum}.} with stress tensor
\be\label{tdust}
T^{\mu\nu}_{\rm{matter}} = \rho \, u^\mu u^\nu \, ,
\ee
where $\rho$ is the energy density and $u^\mu$ is the velocity field. Take the metric ansatz
\be\label{vvmet}
\d s^2 = -F(r,v)\d v^2 + 2 \d v \d r + r^2 \d\varphi^2 ~.
\ee
The coordinate $v$ parameterizes an ingoing null direction and the boundary is located at $r\rightarrow \infty$. In the bulk, $v$ is an ingoing null coordinate, but at the boundary $v$ is identified with ordinary Lorentzian time $t$ in the dual CFT. The energy momentum tensor of an infalling thin shell of null dust is then 
\be\label{tvaidya}
T_{\mu\nu}^{\rm matter} = \frac{8G_{\rm N}\,m+1}{16\pi G_{\rm N}} \frac{\delta(v-v_0)}{r}\,\delta_{\mu}^v\delta_\nu^v ,
\ee
where $m$ is the mass of the final black hole. In fact, we require $m >0$, as going below this bound would correspond to a conical singularity, rather than a black hole in the final state. The solution of the Einstein equations with a source given by \eqref{tvaidya} is the Vaidya metric,
\be\label{eq.globalVaidya}
F(r,v)=1+\frac{r^2}{\ell^2} -    \left(1+\frac{r_+^2}{\ell^2}\right) \Theta(v-v_0)\,,
\ee
where we have defined $r_+ = \ell \sqrt{8G_{\rm N}\,m}$. This is the solution for the case where the boundary CFT lives on $ S^1$. One can also unwrap the $\varphi$ coordinate to obtain the metric of Poincar\'e Vaidya, which has conformal boundary $\mathbb{R}^{1,1}$ and thus corresponds to the dual CFT on the line.

In order to construct the CFT dual, the idea is to model the null dust by a large number of individual particles, each of which will be created by a corresponding operator insertion in the CFT. This is illustrated in figure \ref{fig:shellCollapse}. The advantage of this approach is that existing large-$c$ techniques in CFT can be applied to a finite number of operator insertions; we will then take the limit of an infinite number of particles, holding the total energy fixed, to derive the dual to null dust.

On the gravity side, a standard calculation shows that pressureless dust is identical to a large number of particles traveling on geodesics.  
In order to produce the thin-shell Vaidya geometry, with matter stress tensor \eqref{tvaidya}, the individual dust particles should start at the boundary of AdS at time $t=0$, spaced uniformly around the $\varphi$ circle.\footnote{There is a subtlety in how we interpret the order of limits that defines the Vaidya spacetime. In the gravity context, it is most natural to consider $G_N$ as a fixed, small parameter, and take a large number of dust particles $n \to \infty$ with $G_N$ held fixed. With this order of limits, the mass of an individual dust particle $m_{\rm dust}$ must be taken to zero so that the total energy stays finite. However, we will interpret Vaidya as a different order of limits: first $G_N \to 0$, then $n\to\infty$, or in other words $1 \ll n \ll \frac{\ell_{\rm AdS}}{\ell_{\rm Planck}}$. In this limit we can treat the spherical shell as a very large number of massive particles with $m_{\rm dust}\ell_{\rm AdS} \gg 1$, while still holding fixed the total energy.  The limits commute, so either order can be interpreted as the Vaidya geometry on the gravity side, but it is the latter point of view that will be taken in the dual CFT, as discussed in detail in section \ref{ss:limitzoo}. This will allow us to treat the dust operators as heavy insertions in the CFT.}

\subsection{Definition of the Collapse State $|{\cal V}\rangle$}

We will derive properties of the Vaidya geometry by considering the large-$c$ limit of CFT observables in $1+1$ dimensions, meaning that we construct a quantum state $|{\cal V}\rangle$ whose expectation values reproduce those computed in the gravity background (\ref{vvmet}) through holography. We now embark on this CFT calculation by first defining the state $|{\cal V}\rangle$.

We will define the collapse state in radial quantization, which is in Euclidean signature. Later, we will analytically continue to Lorentzian time since our goal is to understand real time dynamics. The gravity discussion above motivates the following construction. Denote by $z$ the complex coordinate of the CFT in radial quantization, so that states of the Euclidean CFT on $S^1$ are defined on the circle $|z| = 1$.  For each dust particle located at $z=e_{k}$, with $k=1,\ldots ,n$, we roughly need to insert a primary scalar operator $\psi(e_k,\bar e_k)$ with conformal weight $h_\psi$ (and $\bar h_\psi = h_\psi$) on the unit circle. Such a state, however, is not normalizable, and so we regulate it by inserting the operators instead on the circle $|z| = 1 - \sigma$ for some $\sigma >0$, eventually taken to be small. Distributing the $n$ operators uniformly on the circle, a natural guess for the collapse state is then
\be\label{vdef}
| {\cal V} \rangle =\lim_{n\rightarrow \infty} \frac{1}{\mathcal{N}_{n}} \prod_{k=1}^n \psi(e_k,\bar e_k) | 0 \rangle\, \ , \qquad e_k = (1-\sigma) e^{2\pi i (k-1)/n} \ ,
\ee
where $\mathcal{N}_{n}$ is a normalization and $| 0 \rangle$ is the conformal vacuum. The limiting procedure that defines this somewhat formal expression will be described in detail. We can then compute the expectation values of arbitrary local operators ${\cal Q}$ using 
\begin{multline}\label{eq.expectationValueVaidya}
\langle {\cal V} | {\cal Q}_1(z_1,\bar z_1) \cdots {\cal Q}_p(z_p,\bar z_p) | {\cal V}\rangle =\\ \lim_{n\rightarrow \infty} \frac{1}{|\mathcal{N}_{n}|^2} \left\langle \left(\prod_{i=1}^n e_i^{-2\bar h_\psi}\bar e_i^{-2 h_\psi} \psi (\bar e_i^{-1}, e_i^{-1})\right) \, {\cal Q}_1(z_1,\bar z_1) \cdots {\cal Q}_p(z_p,\bar z_p) \, \left(\prod_{k=1}^n\psi(e_k, \bar e_k)\right) \right\rangle\,,
\end{multline}
where the expectation value on the right-hand side is taken in the vaccum. We will take the scaling dimension of the `probe operators' ${\cal Q}_i$ to be $h_{i},\bar h_{i} \ll c$, so on the gravity side, these insertions do not backreact on the geometry.

A few comments are in order. In radial quantization, the conjugate of a real operator is defined as ${\cal O}(z,\bar z)^\dagger  = z^{-2\bar h}\bar z^{-2h}{\cal O}(\bar z^{-1},z^{-1})$, which to leading order in $\sigma$ results in an operator inserted at the same phase angle but on the circle of radius $1+\sigma$. Primary states in radial quantization are defined by inserting a primary operator at the origin. However, this is not what we want, since primaries are energy eigenstates on the cylinder, with trivial dynamics. The state $|\V\rangle$ has operators inserted elsewhere, so it is clearly not primary and will have true dynamics.

Evidently the expression (\ref{eq.expectationValueVaidya}) instructs us to find an `infinite-point' correlation function. This sounds daunting, but the main technical result of our paper is that a correlation function of the type (\ref{eq.expectationValueVaidya}) with $n\rightarrow \infty$ becomes easy to calculate at large central charge. Although the derivation of the prescription is somewhat technical,  the actual calculations, technique in hand, turn out to be efficient and simple ---  easier than the gravity calculations that we will reproduce.

We will choose the total energy above the threshold where black holes form, rather than conical defects. For an explicit comparison between CFT and gravity data, the reader should consult section \ref{sec.MatchingParams}.

One may naturally ask why we are defining the collapse state by inserting operators in Euclidean time rather than by adding a source to the CFT.  We discuss the equivalence between these two pictures in more detail in appendix \ref{ap:sourcevstate}.

\section{CFT Technology}\label{s:cfttech}

\subsection{Conformal block expansion}
We will compute the probe correlators defined in \eqref{eq.expectationValueVaidya} using the conformal block expansion, as formulated for holographic theories in \cite{Hartman:2013mia}.  In principle, this means iteratively applying the OPE between pairs of operators, until left with a product of 3-point coefficients $c_{ijk}$.  There are many ways to take this OPE, but in the end, crossing symmetry requires any channel to produce the same correlator.

For concreteness, consider the 2-point function of identical probe operators,
\be
G_2 (z_{1},z_{2}) = \langle \V| {\cal Q}(z_1, \bz_1) {\cal Q}(z_2, \bz_2)|\V\rangle \, .
\ee
(The results readily generalize to any even number of probes.) We choose to expand in the channel summarized by the diagram
\begin{equation}\label{exactOPE}
G_2 = 
\sum_{i,j,\dots}
\begin{gathered}
\includegraphics[width=300px]{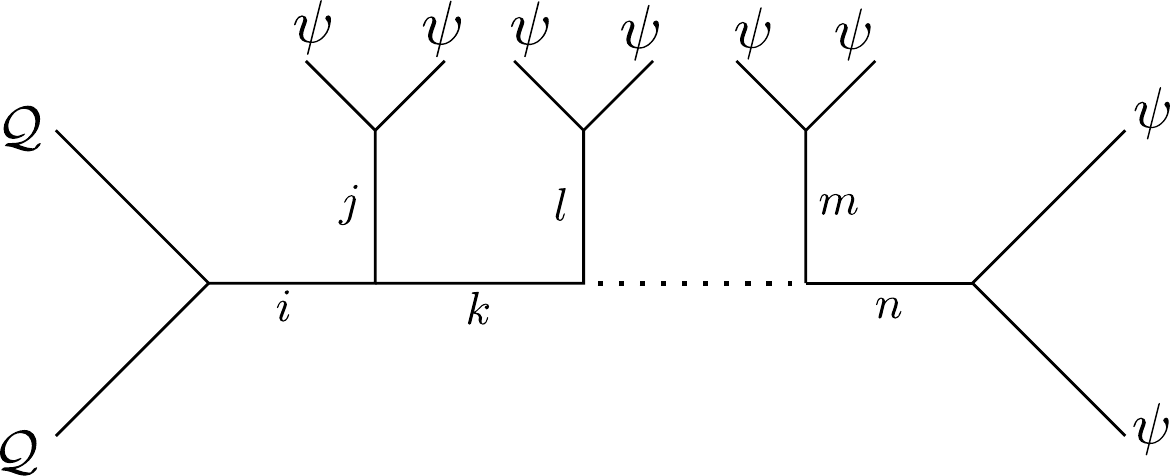}
\end{gathered}
\end{equation}
That is, we first contract the two probes with each other, and each dust operator $\psi(e_k)$ with its conjugate $\psi(e_k)^\dagger$, then contract the resulting operators as indicated.  The internal indices $i,j,k,\dots$ run over all of the primary operators in the CFT.  More explicitly, this diagram stands for the expansion
\be\label{fsum}
G_2(z_{1},z_{2}) = \sum_{i,j,\dots} a_{ijk\dots} \mathcal{F}_{ijk\dots}(z_1, z_2) \overline{\mathcal{F}}_{ijk\dots}(\bz_1, \bz_2) \ ,
\ee
where the function $\mathcal{F}_{ijk\dots}$ is the appropriate Virasoro conformal block, and the constant is the product of OPE coefficients, $a_{ijk\dots}=c_{{\cal Q}{\cal Q}i}c_{\psi\psi j}\cdots$.  The blocks encode the position dependence of the correlator, and are entirely fixed by the Virasoro algebra. Though not written explicitly, they also depend on the choice of channel, the central charge, the internal weights $h_i, h_j, \dots$, the external weights $h_\psi$ and $h_{\cal Q}$, and the insertion points of the dust operators, $e_k$ in \eqref{vdef}. Conformal invariance could be used to fix three of the operator positions, customarily to $0,1,\infty$, but it will be more convenient to leave them as written. A similar computation arises if one is interested in the entanglement entropy of a number of disjoint intervals in heavy eigenstates \cite{Banerjee:2016qca}. In this case one needs to determine a correlator involving two heavy and an arbitrary number of light operators and the dominant channel is given by pairwise fusion of the light operators.

\begin{figure}
\centering
\includegraphics{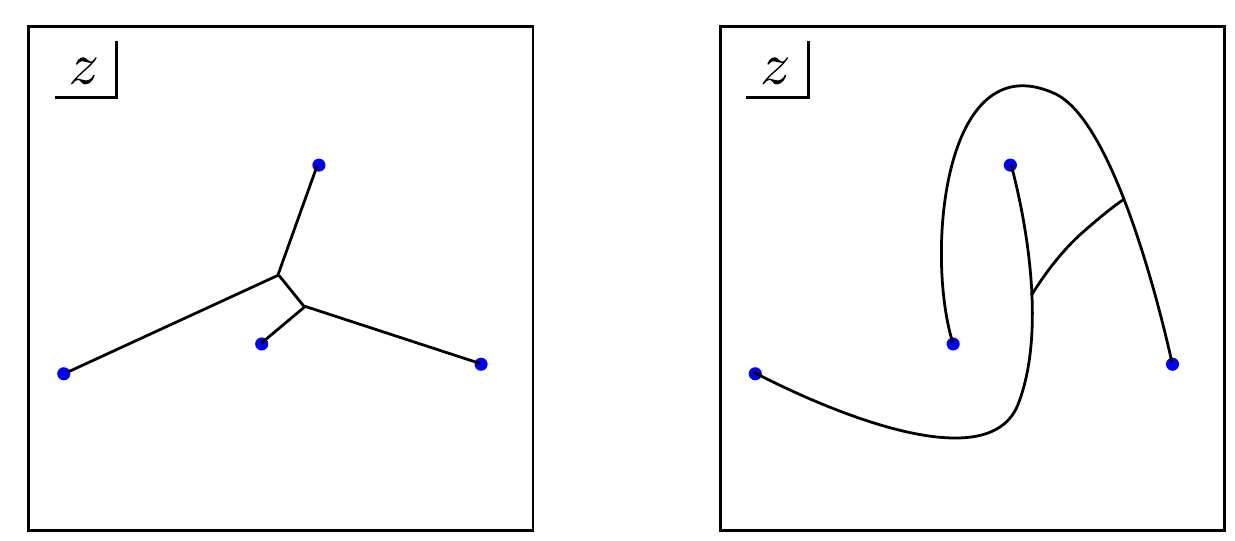}
\caption{\small Two different OPE channels for a given four-point function. These two channels have the same trivalent graph, but correspond to two distinct conformal block expansions. They differ by moving one insertion point around another.\label{f:twoblocks}}
\end{figure}

Note that the diagram \eqref{exactOPE} does \textit{not} uniquely specify the OPE channel.  To specify it uniquely, we must say not only which operators are contracted, but also the set of paths $\Gamma$ used to bring these operators together on the complex plane.  For example, figure \ref{f:twoblocks} depicts two distinct OPE channels for a four-point function. These two possibilities correspond to two distinct sums over conformal blocks, so we will refer to them as different channels.

\subsection{It from Id}

In complete generality, it is impossible to say more. The spectrum of primaries and the OPE coefficients depend on the microscopics of the CFT, so at this point we need to specialize to a class of CFTs that can be expected to have holographic duals.  To this end, we now state the main technical assumption of this paper: \\

{\it In the OPE channel \eqref{exactOPE}, the dominant contribution at large $c$ comes from the identity Virasoro block, that is the unit operator $\mathbf{1}$ and all its Virasoro descendants running on the internal lines: $T, \partial T, T^2, T\partial T$, etc. This continues to apply in the limit $n\rightarrow \infty$.}\\

 This should be viewed as a statement about the type of CFT which admits a large-$c$ limit with emergent gravity.  For certain correlators in a special class of CFTs, it can be derived from first principles \cite{Hartman:2013mia,Hartman:2014oaa}, but we will not restrict to such cases, leaving open the question of exactly what class of theories is captured by this approximation.  Roughly speaking, these are theories with large $c$ and a sufficiently sparse spectrum of low dimension operators. This is motivated by the observation that in the large-$c$ limit, the Virasoro block for heavy external operators exponentiates as \cite{zamolodchikov1986two,zamo}
 \be
 \mathcal{F} \approx e^{-\frac{c}{6}f} \ ,
 \ee
 where $f$ depends on the internal and external conformal weights and the central charge only in the ratio $h/c$. The sum over conformal blocks \eqref{fsum} is then a sum of exponentials, and by the usual saddlepoint logic, we expect this sum to be well approximated by the largest term.  If there are very few primaries of low dimension, then this is the one with the strongest singularity as the operators come together, which is the identity block, denoted
 \be\label{idexp}
  \mathcal{F}_0 \approx e^{-\frac{c}{6}f_0} \ .
\ee
This block encodes the contribution of the unit operator and all of its descendants.
Other light operator exchanges can give comparable contributions to the correlator, but since these have $h/c \to 0$, they have the same conformal block in the large-$c$ limit, and so affect only  the coefficient of $e^{-\frac{c}{6}f_0}$ which is subleading at large $c$.

The assumption that a given OPE channel is dominated by the identity block can only hold within some finite range of kinematics --- it cannot hold for arbitrary positions of the operators.  This would violate crossing symmetry, since the identity in a given channel does not account for the identity in a different channel or vice-versa. This means that as we vary the kinematics, we expect `phase transitions' where the identity operator and the heavy operators in a given channel exchange dominance \cite{Headrick:2010zt}.  The minimal possibility consistent with crossing symmetry is that the exact correlator is approximated by the identity contribution in whichever channel is largest.  This is exactly what was proved for the torus partition function in \cite{Hartman:2014oaa}, and we will assume the same applies to the correlators considered here.

In summary, we assume 
\be\label{gtf}
G_2 (z_{1},z_{2}) \approx \max_{\Gamma} \exp\left[-\frac{c}{6}f_0(z_1,z_2) -\frac{c}{6} \bar{f}_0(\bz_1, \bz_2)\right] \ .
\ee
The maximum is taken over channels $\Gamma$ of the type \eqref{exactOPE}.  All of these channels have the same trivalent graph, but as discussed above, differ in the paths used to define the OPE.

This gives universal results for 2D CFTs with sparse spectrum at large central charge, which translate to a universal sector of quantum gravity theories in AdS$_3$ in the semiclassical limit. This approximation will reproduce the gravity answer, but cannot be the full story from the CFT point of view, and indeed we will argue that it must break down at late times.

\subsection{A zoo of limits}\label{ss:limitzoo}
Before proceeding to the calculation of the Virasoro vacuum block, we pause to clarify the various limits involved in the definition of the collapse state $|\V\rangle$ and the probe correlators that we aim to compute. A variety of limits are needed: 
\begin{itemize}
\item $c\to \infty$, the holographic limit;
\item $n \to \infty$ to produce a smooth matter distribution from the discrete dust particles;
\item $\sigma \to 0$, so that the spherical shell of matter starts exactly from the boundary at time $t=0$;
\item $h_{\psi}/c \to 0$, in order to keep the energy  $E\sim h_{\psi}n/\sigma$ of order $c$ in the above limits;
\item and $h_{\cal Q} \to \infty$, since we intend to compare the CFT correlators to the geodesic (WKB) approximation on the gravity side (but $h_{\cal Q}/c \to 0$ so that we can ignore the backreaction of the probe particles). 
\end{itemize}
How to define the precise order of limits is guided by two considerations: applicability of the exponentiated formula for the Virasoro conformal block, and ensuring that $E/c \sim h_{\psi}n/(c\sigma)$ is fixed in the limit in order to agree with the finite mass black hole.  With some foresight, the limit we will take to compute the leading large-$c$ dependence of the correlator $G_2$ is
\be\label{approxe}
G_2 \approx \exp\left( c \lim_{n \to \infty} \lim_{c \to \infty} \frac{1}{c}\log \mathcal{F}_0 \overline{\mathcal{F}}_0\right)
\ee
where we scale
\be\label{approxeb}
h_{\cal Q} \sim \varepsilon c , \quad h_{\psi} \sim \frac{\varepsilon}{n}c
\ee
for some fixed $\varepsilon \ll 1 $. This can be done at finite $\sigma$, but for comparison to Vaidya, we are interested in $E\sim c$ (as in the black hole) so choose $\sigma \sim \varepsilon $. All of the final results of the paper, such as the eventual matching of CFT correlators to geodesic lengths, should be understood in the sense of equations (\ref{approxe}-\ref{approxeb}).

\subsection{Semiclassical Conformal Blocks and The Monodromy}

Our task is to compute the large-$c$ Virasoro identity block with $n\to \infty$ dust operator insertions and two (or more) probe insertions. For any finite $n$, the large-$c$ block can be computed, at least in principle, using a monodromy method introduced by Zamolodchikov \cite{zamolodchikov1986two,zamo} (and reviewed in \cite{Harlow:2011ny,Hartman:2013mia}). We first state the general procedure to compute the vacuum block, then describe how to implement it when the operators ${\cal Q}$ are light compared to the combined effect of the operators $\psi$ defining the state.

The monodromy method was originally stated for heavy operator exchange in a four-point function \cite{zamolodchikov1986two,zamo}. It can be extended to heavy operator exchange in higher-point functions, but we will consider only identity exchange, where the procedure is simpler.  Despite the fact that the identity is a light operator, the method still applies with no significant changes \cite{Hartman:2013mia}. Perturbation theory of the monodromy equation, used to compute probe correlators, was introduced in \cite{Fitzpatrick:2014vua}.

\subsubsection{The General Procedure}

\begin{figure}[t!]
\begin{center}
$\Gamma_1$\includegraphics[width=0.38\textwidth]{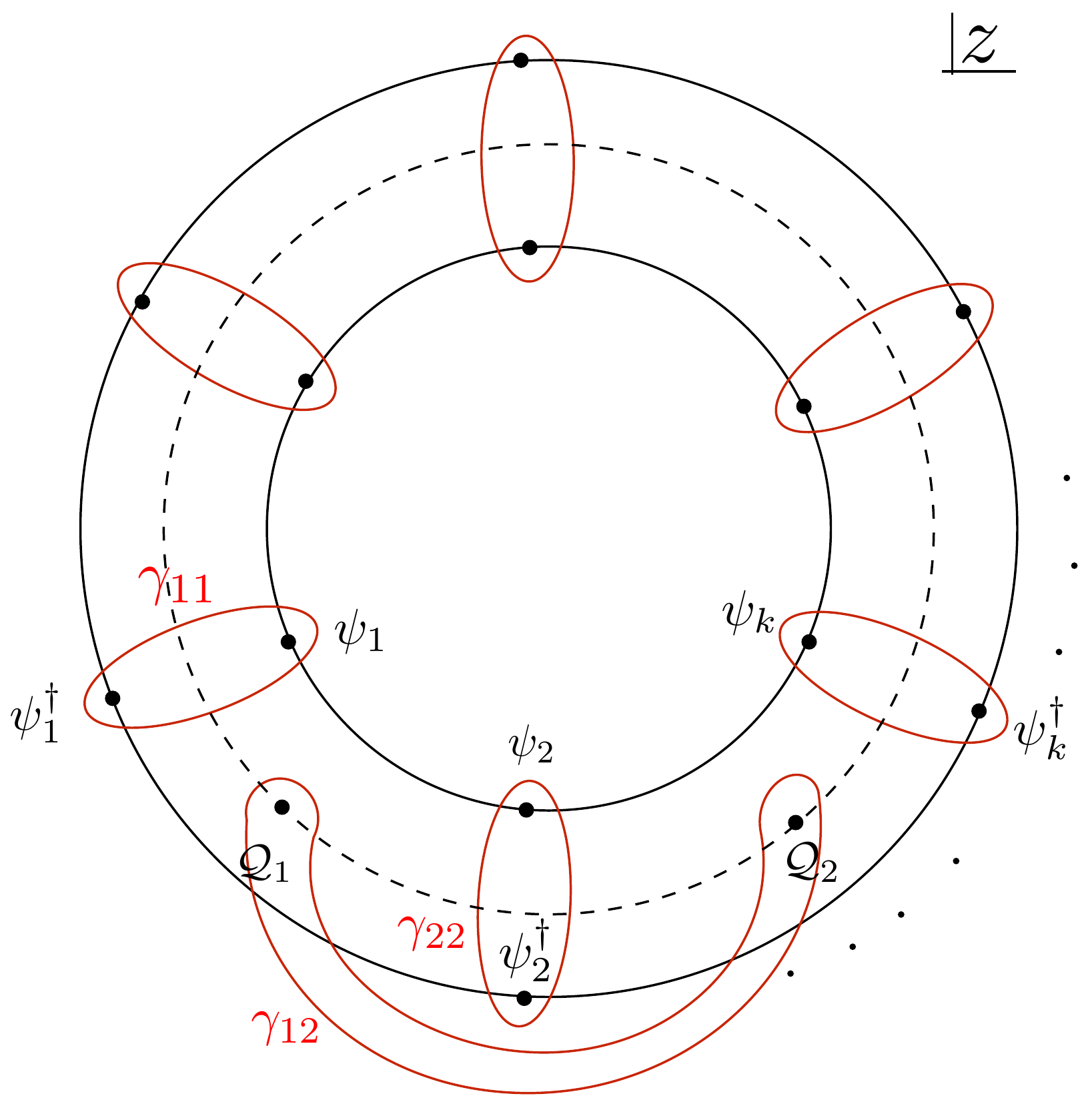}\hskip3em $\Gamma_2$\includegraphics[width=0.38\textwidth]{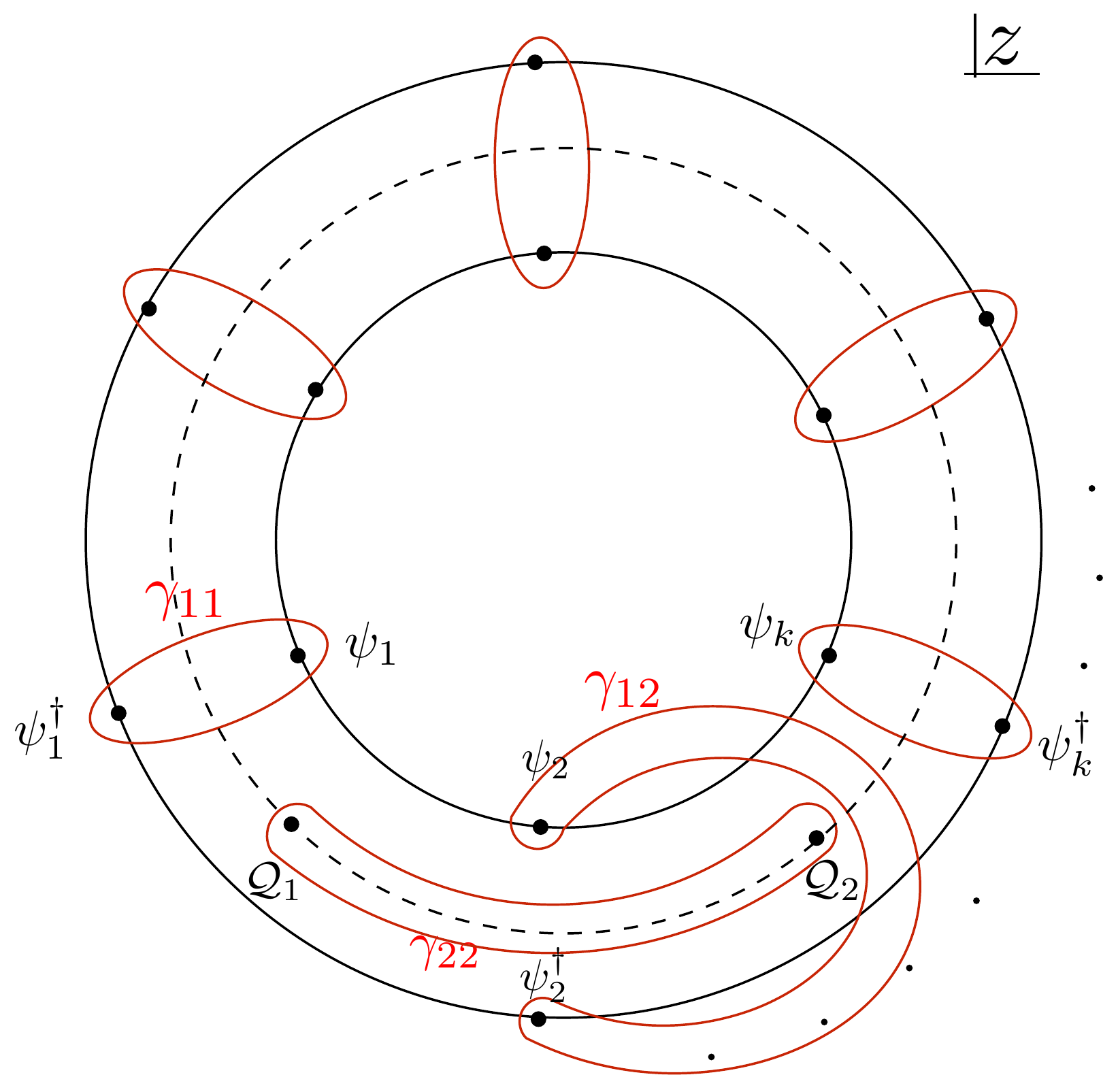}
\caption{\small Two different OPE channels contributing to the correlator (\ref{eq.expectationValueVaidya}). The differential equation (\ref{eq.monodromicODE}) is required to have trivial monodromy around each cycle indicated in red.  \label{fig:OPEchannels} The dashed circle is at $|z|=1$.}
\end{center}
\end{figure}

For a general Euclidean correlator of heavy operators $\langle {\cal O}_1(z_1)\cdots {\cal O}_{m}(z_m)\rangle$, with $m$ finite, the monodromy method to compute the large-$c$ vacuum block is as follows:

\begin{enumerate}
\item Consider the following differential equation on the complex plane:
\be\label{eq.monodromicODE}
\chi''(z) + T_{\rm cl}(z) \chi(z) =0\,,
\ee
where 
\be\label{tclf}
T_{\rm cl}(z) = \sum_{k=1}^m \left( \frac{6 h_k/c}{(z-z_k)^2}  - \frac{c_k}{z-z_k}\right)
\ee
and $h_k$ is the conformal weight of operator ${\cal O}_k$. 
The numbers $c_k$ are called accessory parameters and will be fixed below.

\item A channel is defined by contracting the external operators in pairs, ${\cal O}_k(z_k){\cal O}_l(z_l) \to \bf{1}$.\footnote{This pairing completely determines the vacuum block, but this would not be the case for a conformal block involving the exchange of non-vacuum primaries.  For general operator exchange, we would also need to specify how these exchange operators themselves are paired, and so on.  This is not necessary for the vacuum block because after pairing the external operators, we have a correlator made entirely of stress tensors. Such correlators are fixed by the Virasoro algebra and are independent of the fusion channel. In other words, we are exploiting the fact that
\begin{equation}
\includegraphics{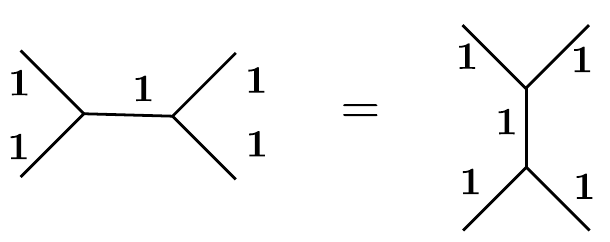}
\end{equation}
viewed as a subgraph inside any OPE diagram.}
 To contract two operators to the identity representation, they must have the same scaling weight, $h_k=h_l$. These contractions are indicated in the complex plane by drawing  non-intersecting closed contours around pairs of operator insertions. We denote the set of all such cycles defining a given channel as $\Gamma$. Two examples of different channels are illustrated in Fig. \ref{fig:OPEchannels}. 

\item The second order differential equation \eqref{eq.monodromicODE} has two independent solutions, say $\chi_1$ and $\chi_2$. These solutions may undergo a monodromy as we follow them along a closed loop $\gamma$ around singular points of the differential equation,
\be
\left(\! \begin{array}{c} \chi_1 \\ \chi_2 \end{array}\!\right)
\to
M_{\gamma} \left(\! \begin{array}{c} \chi_1 \\ \chi_2 \end{array}\!\right) \, ,
\ee
where $M_{\gamma}$ is a two-by-two invertible complex matrix.
The accessory parameters $c_k$ are fixed (as a function of $c$ as well as the $h_k$ and $z_k$) by demanding that the monodromy matrix around each cycle $\gamma \in \Gamma$ is trivial, 
\be
M_{\gamma} = \id~.
\ee
\item The semi-classical conformal block in a given channel $\Gamma$ is determined by integrating the partial differential equations
\be\label{blockexp}
\frac{\partial f_0(z_{1},\ldots ,z_{m})}{\partial z_k} = c_k \ ,
\ee
subject to the boundary condition that $f_0$ has the correct singularity near coincident points. The leading singularity as $z_k \to z_l$ is $(z_l-z_k)^{-2h_k}$, so comparing to \eqref{idexp}, this boundary condition is
\be
f_0(z_{1},\ldots , z_{m}) \simeq \frac{12h_k}{c}\log(z_l-z_k) \quad \mbox{as} \quad z_l \to z_k \ .
\ee

\end{enumerate}

\noindent As a byproduct, this method also computes for us the expectation value of the CFT stress tensor.  In the case that the correlator is dominated by the vacuum block, the relation is simply
\be\label{tclval}
\langle T(w) {\cal O}_1(z_1) {\cal O}_2(z_2) \cdots {\cal O}_m(z_m)\rangle = \left[\frac{c}{6}T_{\rm cl}(w)+O(c^0)\right]\langle {\cal O}_1(z_1) \cdots {\cal O}_m(z_m)\rangle \ .
\ee
The necessity of the factor $c/6$ is apparent from the coefficient of the leading singularities in \eqref{tclf}, which is $h_{k}$ for the usual normalization of the CFT stress tensor.

\subsubsection{Heavy-Light Perturbation Theory}\label{sec:HeavyLight}
To compute correlators for probe operators of equal weight $h_{\cal Q}$ in the collapse state $|\V\rangle$, the relevant differential equation has the following $T_{\rm cl}$:
\be\label{eq.finitenStressTensor}
T_{\rm cl}(z) = \sum_{k=1}^{2n} \left( \frac{6h_\psi /c}{(z-x_k)^2} - \frac{c_k}{z- x_k} \right) + \sum_{k=1}^{N_{\cal Q}}\left(\frac{6h_{\cal Q}/c}{(z-z_k)^2} - \frac{c^{\cal Q}_k}{z-z_k}  \right)\,.
\ee
Here we have split up $T_{\rm cl}(z)$ into the contributions from the insertions $\psi$ defining the state, and from the probe insertions ${\cal Q}$. The singular points $x_k$ are taken to be $x_1, \ldots ,x_n = e_1,\ldots, e_n$ and $x_{n+1},\ldots, x_{2n} = 1/\bar e_1,\ldots, 1/\bar e_n$, while at this stage $z_{1},\ldots, z_{k}$ are left arbitrary.

Suppose that the second contribution in (\ref{eq.finitenStressTensor}) is parametrically smaller than the first. Later on, we will choose the dimension of the dust operators to scale so that the first term is $O(c^0)$, so we should choose the probes to have scaling dimension $h_{\cal Q} \sim \varepsilon c$ with $\varepsilon \ll 1$. In other words, $1 \ll h_{\cal Q} \ll c$.  In gravity-inspired language we view the $\psi$'s as creating a {\it background} which is {\it probed} by the ${\cal Q}$'s. A method to solve the monodromy problem in this limit, with a finite number $n$ of background insertions, was introduced in \cite{Fitzpatrick:2014vua}. We split up the energy momentum tensor into a heavy background contribution and a light probe contribution
\be\label{eq.perturbativeStressTensor}
	T_{\rm cl} = T_{\rm H} + \varepsilon\, T_{\rm L}\, ,
\ee
corresponding to the two terms in \eqref{eq.finitenStressTensor}.
We then want to solve the differential equation (\ref{eq.monodromicODE}) perturbatively in $\varepsilon$. Let us define
\be
\chi = v+ \varepsilon w\,.
\ee
Working to first non-trivial order in $\varepsilon$, the differential equation yields
\bea
v'' + T_{\rm H}v &=&0\, ,\label{eq0} \\
w'' + T_{\rm H}w &=& -T_{\rm L}v\, . \label{eq1}
\eea
Let $V = \left( v_1,v_2 \right)^t$ denote a two-vector of linearly independent solutions of \eqref{eq0}. Then the solution at ${\cal O}(\varepsilon)$  can be determined by the method of variation of parameters, and is given by
\be\label{chiz}
\chi(z) = \left( 1 + \varepsilon \int^z_{z_0} \d z' F(z') \right)V(z)\,,
\ee
where $F$ is a $2\times 2$ matrix with components
\be\label{deff}
F_i{}^j = \frac{  v_i \epsilon^{jk} v_k}{v_1 v_2' - v_2 v_1'} T_{\rm L}  
\ee
with $\epsilon^{12}=1$. The lower limit of integration in \eqref{chiz}, $z_0$, is an arbitrary complex number --- we can choose whatever starting point is convenient, and this defines the basis of solutions. The basis also depends on a choice of path in the complex plane, implicit in \eqref{chiz}.

A very nice feature of \eqref{chiz} is that we can compute first-order monodromies with minimal effort.  Suppose we are interested in the monodromy of $\chi$ around one of the probe insertions, $z_k$.  The zeroth order solutions $V$ have no monodromy around this point, so only the $\int F$ term in \eqref{chiz} can contribute. If we start with the solution $\chi(z)$ in a neighborhood of the point $z_0$, and analytically continue this solution along a closed curve that encircles the singularity and returns to a neighborhood of $z_0$, then after traversing this loop the solution is
\be\label{eq:loop}
\chi \to \left(1 + \varepsilon \oint_{z_i} F  + \varepsilon \int_{z_0}^z F\right)V \ .
\ee
In regions where $T_{\rm H}$ and $T_{\rm L}$ are both meromorphic, so is $F$, and the first integral in \eqref{eq:loop} gives a residue. So in these regions, the monodromy matrix on this path is
\be\label{eq:loopmonodromy}
M_{z_i} =1 + 2\pi i \varepsilon \, \mbox{Res}_{z_i}F \ .
\ee
This technique was applied to a meromorphic stress tensor in \cite{Fitzpatrick:2014vua}, and the residues completely fix the conformal block. As we will see, our situation is more complicated, since $T_{\rm H}$ is not meromorphic globally, and the calculation will in general require more than just residues.

\subsection{Continuum Monodromy Method}
\subsubsection{The Stress Tensor at ${\cal O}(\varepsilon^0)$}\label{sec.StressAtOrderZero}
In the limit $n\rightarrow \infty$ there is an infinite number of $\psi$ insertions and only a finite number of light operators.
At ${\cal O}(\varepsilon^0)$ we thus want to solve the monodromy problem for an infinite number of operators. We now describe how to tackle this limit directly, leading to a drastic simplification of the calculation. We start by writing the stress tensor\footnote{In the present case the continuum stress tensor can be arrived at by taking the continuum limit of Eq. (\ref{eq.finitenStressTensor}), so that $e_k = (1+\sigma)e^{\frac{i2\pi j}{n}}\rightarrow e(\theta) = (1+\sigma)e^{i\theta}$, and $\sum_k \rightarrow \frac{n}{2\pi}\int d\theta$. However, the method applies much more widely, resulting in the general continuum expression (\ref{eq.ContinuumStress}).} in the limit $n\rightarrow \infty$ as
\be\label{eq.ContinuumStress}
T_{\rm H}(z,\bar z) = \int \d^2 w\, s(w,\bar w)\left[\frac{6\hat{h}_\psi/c}{(z-w)^2} - \frac{c[s,w,\bar w]}{z-w}  \right]
\ee
where $s(w,\bar w)$ is a weighting function for the source insertions, and $c[s;w; \bar w]$ is an `accessory functional.' The normalized weight is $\hat{h}_\psi = n h_{\psi}$, which is held fixed as $n \to \infty$. Note that $s(w, \bar w)$ could be traded for a space-dependent scaling dimension, $\hat{h}(w,\bar w)$, so all that really matters is the scaling weight density. The form of this stress tensor can be derived by the usual limiting procedures from the sum (\ref{eq.finitenStressTensor}) and represents the same limit described in section \ref{ss:limitzoo}. Although (\ref{eq.ContinuumStress}) looks formally like a holomorphic function of $z$, this is not the case; it has
non-holomorphic dependence on the source location $w,\bar w$ and after performing the integral this will introduce a manifest dependence on $\bar z$. In particular, $\bar\partial T \neq 0$, and the non-holomorphicity is not limited to isolated points as it would be for a meromorphic stress tensor. (This qualitatively new feature is what prevents us from adopting the simplified approach to heavy-light blocks developed in \cite{Fitzpatrick:2015zha}.)

For expectation values in global Vaidya (\ref{eq.expectationValueVaidya}) we take $s$ to have support on the two shells of radius $1+\sigma$ and $1-\sigma$ where operators are inserted. More specifically, we choose
\be
s(w,\bar w) = \delta \left( |w| - 1 - \sigma\right) +  \delta \left( |w| - 1 + \sigma \right)\,.
\ee
Splitting into the inner and outer shells, 
\be\label{eq.stressOnCircle}
T_{\rm H} = \int_{0}^{2\pi} \frac{\d\theta}{2\pi} \left[ \frac{6\hat{h}_\psi/c}{(z-(1 + \sigma)e^{i\theta})^2} +  \frac{6\hat{h}_\psi/c}{(z-(1 - \sigma)e^{i\theta})^2} - \frac{c_{+}(\theta)}{z-(1 + \sigma)e^{i\theta}} - \frac{c_{-}(\theta)}{z-(1 - \sigma)e^{i\theta}}\right]\,.
\ee
Now we need to fix the accessory functions $c_{\pm}$. Our task is to implement the continuum version of the channel depicted in figure \ref{fig:OPEchannels}, where each $\psi$ is contracted with its conjugate. In the continuum limit, this channel has a rotational symmetry which can be used to fix $c_{\pm}(\theta)$ up to overall coefficients:\footnote{In cylinder coordinates $z=e^{w}$, the residues should be independent of the angle Im $w$. Translating $T_{ww} \sim \cdots +  \frac{K}{w-w_i}+\cdots $ to the plane using $T_{zz} = \frac{1}{z^2}T_{ww} + \cdots$ gives the residue $\frac{K z_i^{-1}}{z-z_i}$. The factor of $z_i^{-1}$ is the origin of the $e^{-i\theta}$ in \eqref{accz}. The first equality $c^+=-c^-$ comes from imposing regularity of the stress tensor at infinity.}
\be\label{accz}
c_{+}(\theta) = - c_{-}(\theta) = K\, e^{-i\theta} \ ,
\ee
where $K$ is a constant that will shortly be fixed. This allows us to rewrite $T_{\rm H}$ in terms of a differential operator acting on a simpler integral, namely
\be\label{eq.SimpleIntegral}
T_{\rm H}(z,\bar z)=\left[\frac{6\hat{h}_{\psi}}{c} \frac{\partial}{\partial \sigma} - K  \right]
\int_{0}^{2\pi} \frac{\d\theta}{2\pi} e^{-i\theta} \left(\frac{1}{z-(1 + \sigma)e^{i\theta}} - \frac{1}{z-(1 - \sigma)e^{i\theta}}   \right) \ .
\ee
The remaining integral evaluates to zero for $|z| < 1-\sigma$, to $-\frac{1}{z^2}(1-\sigma)$ in the annulus $1-\sigma < |z| < 1 + \sigma$, and to $\frac{2 \sigma}{z^2}$ for $|z| > 1+\sigma$. Acting with the differential operator on these expressions gives $T_{\rm H}$. Regularity at infinity requires $T_{\rm H} \sim z^{-4}$, which sets
\be
K = \frac{6\hat{h}_\psi}{c\sigma} \, \cdot
\ee
The contributions of the derivative in from of (\ref{eq.SimpleIntegral}) as well as additional delta-function contributions to the integral from $|z| = 1 \pm \sigma$ are subleading in the Vaidya limit $|\sigma|\ll 1$. Therefore we find for the final answer in this limit
\begin{align}\label{eq.ContinuumStressTensor}
T_{\rm H}(z,\bar z) &= \frac{K}{z^2}\Theta\left(  |z| - 1 + \sigma \right) \Theta \left(1- |z| + \sigma \right)
\end{align}
We have thus found that the stress tensor is piecewise holomorphic.
 The dependence on $|z|$ spoils the holomorphicity of the stress tensor explicitly.

Since the accessory functions $c_{\pm}$ were completely fixed by symmetries and regularity, what we have just constructed must be the continuum limit of the channel where each $\psi$ is contracted with its conjugate, as in figure \ref{fig:OPEchannels}. This will be confirmed below by explicit calculation of the monodromies.

\subsubsection{Matching parameters to the gravity side}\label{sec.MatchingParams}
The heavy stress tensor \eqref{eq.ContinuumStressTensor} is simply a constant supported on a narrow annulus around the unit circle where $T_{\rm H}(z) = K/z^{2}$. The total dimensionless energy $E$ associated to this stress tensor is
\be
E =2   \left(\frac{cK}{6} - \frac{c}{24}\right) \, .
\ee
The factor of 2 comes from adding the anti-holomorphic contribution (since everything we have done applies also to $\overline{T}_{\! \rm H}$), the factor $\frac{c}{6}$ in the first term comes from the relative normalization of $T_{\rm cl}$ and $\langle T\rangle$ (see \eqref{tclval}), and the shift by $-\frac{c}{24}$ is the usual Casimir energy in going from the plane to the cylinder.

On the gravity side, the total energy is the mass $m$ of the black hole, and the central charge takes the Brown-Henneaux value $c = \frac{3\ell}{2G_N}$. There is a relative factor of $\ell$ in the usual conventions for CFT energy and bulk energy to account for the units: $E= m\ell$. Therefore the identification of parameters, in order for our state $|\V\rangle$ to produce a black hole of mass $m$, is
\be\label{gravityIdent}
K = 2 m G_{\rm N}  + \frac{1}{4}\, \cdot
\ee
From this we conclude that $K$ must be larger than $1/4$ in order to create a black hole rather than a conical defect.

\subsubsection{The Stress Tensor at ${\cal O}(\varepsilon)$}
Let us now assume that the ${\cal Q}$ insertions are light so that we may take $6 h_{\cal Q}/c = \varepsilon$ as a small parameter. We then have an expression of the form of Eq.\ \eqref{eq.perturbativeStressTensor}, where $T_{\rm H}$ is given by \eqref{eq.ContinuumStressTensor} and 
\be\label{Tlight}
T_{\rm L} (z)= \frac{1}{(z-z_1)^2} + \frac{1}{(z-z_2)^2} - \frac{b_1}{z-z_1} - \frac{b_2}{z-z_2}\, ,
\ee
where the $b_k$ are related to the usual accessory parameters via $b_k\equiv c_k^\mathcal{Q}/\varepsilon$. At this stage we have specialized to $N_{\cal Q}=2$, \textit{i.e.\ }a two-point function in the collapse background, though the method naturally generalizes to any finite number of probe insertions.

\subsection{Solutions of the monodromy equation}

\subsubsection{Solutions at order $\varepsilon^0$}
\begin{figure}[t!]
\begin{center}
\includegraphics[width=0.48\textwidth]{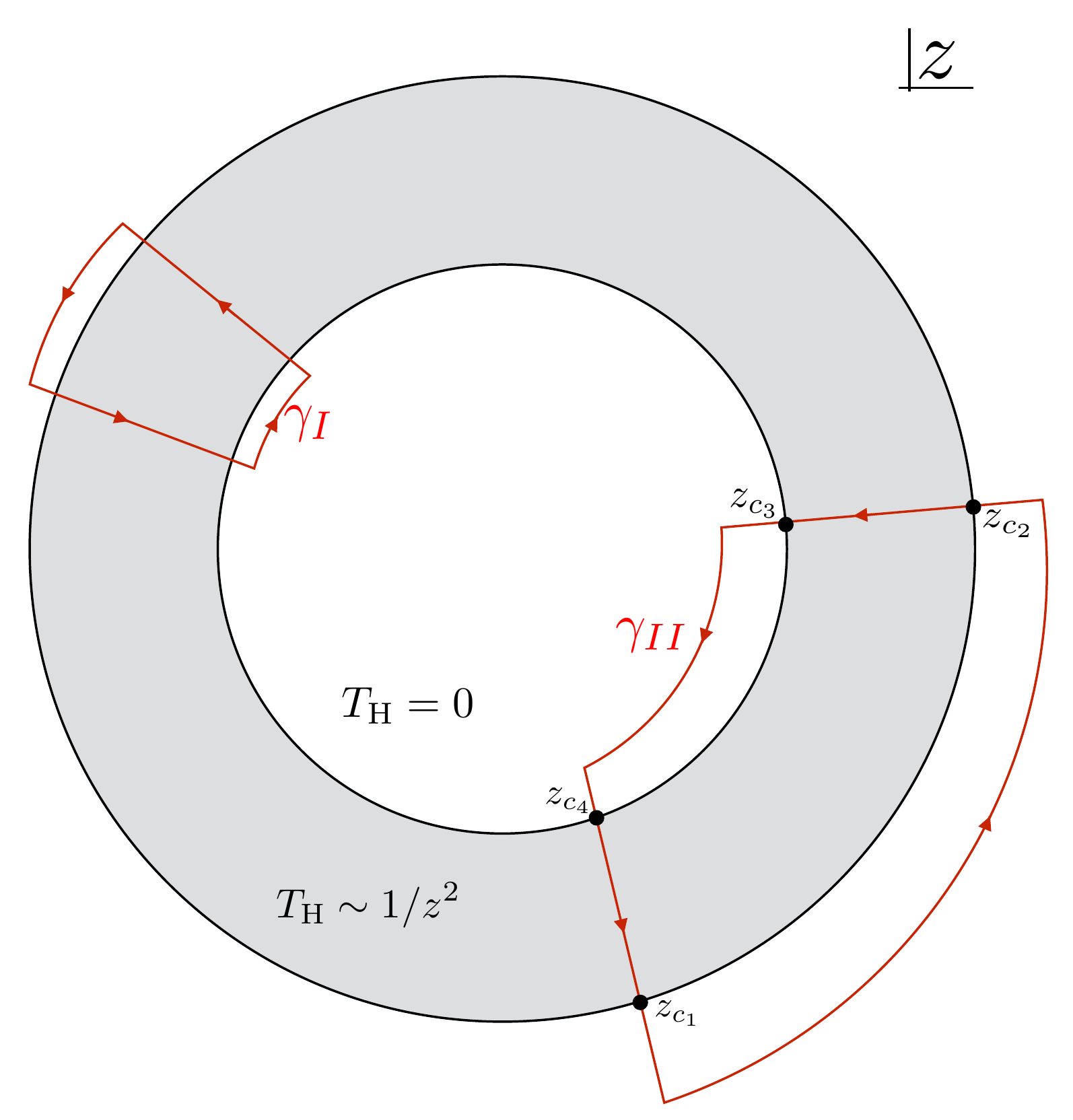}
\caption{\small Two different cycles $\red{\gamma_I}$ and $\red{\gamma_{II}}$ with trivial monodromy for Eq. (\ref{eq.monodromicODE}) using the expression (\ref{eq.ContinuumStressTensor}) for the stress tensor.  In fact any loop straddling the annulus in this fashion has trivial monodromy for the stress tensor (\ref{eq.ContinuumStressTensor}). \label{fig:ContinuumMonodromy}}
\end{center}
\end{figure}

Now that we have the stress tensor, the next step is to solve the differential equations \eqref{eq0}, \eqref{eq1}. The first equation is simple. We choose the basis of solutions inside the annulus
\be\label{basisv}
V (z)= \left( \begin{array}{c}
z^{\frac{1}{2}(1-i\rho)}\\
z^{\frac{1}{2}(1+i\rho )}
\end{array}\right) \ ,
\ee
where 
\be\label{eq:rhodef}
\rho\equiv \sqrt{4K-1} \ .
\ee
(There is a branch cut in \eqref{basisv}, but we will only use this basis locally so this is not a problem.)
Outside the annulus, where $T_{\rm H}(z) = 0$, we choose the basis
\be\label{basisvt}
\tilde{V}(z) = \left( \begin{array}{c}
1\\
z
\end{array}\right) \ .
\ee
We will need to solve \eqref{eq0} along contours that cross from inside to outside the annulus, by matching the solution of the differential equation on both sides. This matching depends on the crossing point $z_c$.  A given solution inside the annulus must match onto some linear combination of our basis solutions outside the annulus, so we can define a matching matrix $J$ that relates the solution $V$ inside to a solution $J \tilde{V}$ outside.  For a general value of $\sigma$, this matching will require solving across the delta function.  This can be done, but to simplify the calculation we assume from here on that
\be
\sigma \ll 1 \ ,
\ee  
which is the Vaidya limit on the gravity side.
Then we see from \eqref{eq.ContinuumStressTensor} that the discontinuity induced by the delta function is subleading, and we can define the matching matrix simply by
\be
V(z_c)  = J_0(z_c) \tilde{V}(z_c) \ , \qquad V'(z_c)  = J_0(z_c) \tilde{V}'(z_c)  \, .
\ee
This yields
\be\label{eq:zerothorderpatch}
J_0(z_c)=\frac{1}{2}\begin{pmatrix}z_c^{\frac{1}{2}(1-i \rho)}(1+i \rho ) &z_c^{-\frac{1}{2}(1+i \rho)}(1-i \rho )\\z_c^{\frac{1}{2}(1+i \rho)}(1-i \rho ) &z_c^{-\frac{1}{2}(1-i \rho)}(1+i\rho )\end{pmatrix} \ .
\ee
Now we can use these solutions to confirm that the heavy stress tensor \eqref{eq.ContinuumStressTensor} indeed corresponds to the continuum limit of the channel where each $\psi$ is contracted with its conjugate, as in figure \ref{fig:OPEchannels}. The differential equation \eqref{eq.monodromicODE} should have trivial monodromy along a path that encloses any number of $\psi$'s and their conjugates. Two examples of such loops are shown in Fig.\ \ref{fig:ContinuumMonodromy}. To compute the monodromy, we need to construct a solution to the differential equation along such a loop. Let us start with the solution $V$ inside the annulus.  Matching to the exterior, $|z| > 1+\sigma$, the solution is $J_0(z_{c_1})\tilde{V}$, where $z_{c_1}$ is the crossing point indicated in the figure. Continuing in this way, we follow the solution all the way around the contour. When we get back to the starting point, the solution is $M V$ with monodromy matrix
\be
M = J_0(z_{c_1})J_0(z_{c_2})^{-1}J_0(z_{c_3})J_0(z_{c_4})^{-1} \ ,
\ee
where (see figure \ref{fig:ContinuumMonodromy})
\be
z_{c_1} = (1+\sigma)e^{i \phi_1} , \quad
z_{c_2} = (1+\sigma)e^{i\phi_2} , \quad
z_{c_3} = (1-\sigma)e^{i\phi_2} , \quad
z_{c_4} = (1+\sigma)e^{i\phi_1} \ .
\ee
Using \eqref{eq:zerothorderpatch} gives trivial monodromy $M = 1_{2\times 2} + \mathcal{O}(\sigma)$, as claimed.

\subsubsection{Solutions at order $\varepsilon$}\label{sec:oepsilon}

Using the notation of subsection \ref{sec:HeavyLight}, the general solution of the order-$\varepsilon$ equation is now provided by \eqref{chiz},
\bea\label{chinout}
\chi_{\rm in}(z) &=& \left(1 + \varepsilon \int^z F\right) V(z) \\
\chi_{\rm out}(z) &=& \left(1 + \varepsilon \int^z \tilde{F}\right)\tilde{V}(z)
\eea
where the matrices $F$ and $\tilde{F}$, defined in \eqref{deff}, are
\begin{align}\label{eq:FFt}
F(z)&=\frac{z\,T_{\rm L}(z)}{i \rho}\begin{pmatrix} 1 &-z^{-i \rho}\\z^{i \rho} &-1\end{pmatrix}~,\quad\quad\tilde{F}(z)=T_{\rm L}(z)\begin{pmatrix} z &-1\\z^2 &-z\end{pmatrix} \ .
\end{align}
To fully specify the solutions in \eqref{chinout} we must also choose a basepoint and path for the integrals. These are chosen in different ways below according to the details of the situation.

\section{Calculation of CFT Correlators}

Now that we've set up the necessary formalism and determined $T_{\rm H}$ to leading order in $\varepsilon$,  we turn to the explicit computation of real-time correlation functions in the state $|\mathcal{V}\rangle$. The procedure is summarized as follows. First, with light probe operators inserted at complex (Euclidean) points $z_1$ and $z_2$, we fix the accessory parameters  $b_{k}$ in $T_{\rm L}$ by demanding that solutions to \eqref{eq.monodromicODE} have trivial monodromy around a given path $\gamma$ encircling both points. We then use (\ref{blockexp}) and (\ref{idexp}) to obtain an expression for the semiclassical identity conformal block, which will depend nontrivially on $\gamma$. The dominant contribution will come from the path that maximizes (minimizes) $\mathcal{F}_0$ ($f_0$). We then analytically continue the insertion points $z_{1,2}$ to Lorentzian times.

We will exhibit this method in two examples of increasing difficulty. The first is the equal-space auto-correlation function $G(t_1,t_2)$ with times taken before and after a global Vaidya quench. We find a simple analytic formula for this correlator in a CFT living on a circle of size $R$. The finite-$R$ result has never been calculated on the gravity side, but taking $R\to \infty$, our CFT result precisely matches a planar Vaidya-AdS$_3$ geodesic length, as computed numerically in \cite{AbajoArrastia:2010yt} and analytically in \cite{Balasubramanian:2012tu}. The second example is the growth of the entanglement entropy $S_{\rm EE}$ of an interval of length $L$ following the Vaidya quench. Our calculations again match the known gravity results \cite{Hubeny:2007xt,Balasubramanian:2011ur,Balasubramanian:2010ce,Ziogas:2015aja}.

\subsection{The equal-space auto-correlation function $G(t_1,t_2)$}
We wish to compute $G(t_1,t_2)$ for $t_1<0<t_2$, where the Vaidya quench occurs at Lorentzian time $t=0$. This correlator probes the physics of thermalization.
The Euclidean correlator of interest thus has probe insertions at points $z_1$ and $z_2$ along the imaginary axis with $z_2$ inside the annulus where $T_{\rm H}=K/z^2$, and $z_1$ positioned outside the annulus, where $T_{\rm H}$ vanishes. The reason for this choice is the following: had we inserted both points below the strip $|z_i|<1-\sigma$, the monodromy prescription would give the vacuum answer (as expected for Lorentzian times $t_1<t_2<0$). Had we instead inserted the probe operators inside the strip $1-\sigma<|z_i|<1+\sigma$, we would simply find the thermal auto-correlation function (as expected for $0<t_1<t_2$). 
\begin{figure}[t]
\begin{center}
\includegraphics[width=0.55\textwidth]{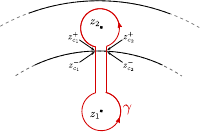}
\caption{\small Path $\gamma$ defining the channel of our correlation function.  The black solid lines are the shockwave insertions at $|z| = 1 \pm \sigma$.  The path $\gamma$ actually crosses the shockwave twice at the same point, but the crossings are separated in the figure for clarity.\label{fig:G}}
\end{center}
\end{figure}

To find the identity block, the first step is to compute the monodromy along the contour in figure \ref{fig:G}. The crossing points in this diagram are actually all equal,
\be
z_c \equiv z_{c_1}^+ = z_{c_1}^- = z_{c_2}^+=z_{c_2}^- \ ,
\ee
but they have been separated in the figure to illustrate how they lie on different points along the contour $\gamma$. We will construct the global solution of the differential equation along this contour, $X = (\chi_1, \chi_2)^t$, following section \ref{sec:oepsilon}.
A basis of solutions inside and outside the annulus is
\bea
\chi_{\rm  in}(z)&=&\left(1+\varepsilon\int_{z_c}^z F\right)V(z)~,\\
\chi_{\rm out}(z)&=&\left(1+\varepsilon\int_{z_c}^z \tilde{F}\right)\tilde{V}(z)~\ .
\eea
In all of the expressions that follow, the integral is taken along a short, topologically trivial path connecting the upper and lower limits of integration.  As discussed in section \ref{sec:HeavyLight}, for a meromorphic stress tensor $T_{\rm cl}$ the monodromy matrix would come directly from the residues of $F$ and $\tilde{F}$ around the singularities $z_i$. Since $T_{\rm cl}$ is not meromorphic, we must include contributions from matching the solution across the annulus. Let us see how this works. 

To construct a global solution along the path $\gamma$, we start at the base point $z_{c_2}^+$. In a neighborhood of this point, choose the solution
\be
X(z) = \chi_{\rm in}(z) \qquad (z \sim z_{c_2}^+) \ .
\ee
Now, follow the path counterclockwise around $z_2$ to get to the first crossing point $z_{c_1}^+$. The solution in a neighborhood of $z_{c_1}^+$ is then
\begin{align}
X(z)&=\left(1+2\pi i \varepsilon\, \text{Res}_{z_2}\,F+\varepsilon\int_{z_c}^{z} F\right)V(z)\nonumber\\
&=\left(1+2\pi i \varepsilon\, \text{Res}_{z_2}\,F\right)\chi_{\rm in}(z)~,
\end{align}
where we have picked up the residue of $F$ by integrating around $z_2$ and have neglected terms of $\mathcal{O}(\varepsilon^2)$ in going from the first to the second line. We now match the solution across $z_{c_1}$. Define the matrix $J(z_{c_1})$ such that $\chi_{\rm in}(z_{c_1}^+)=J(z_{c_1})\chi_{\rm out}(z_{c_1}^-)$. This matching matrix $J(z_c)$ is related to the zeroth order matching matrix $J_0(z_c)$ given in (\ref{eq:zerothorderpatch}) by 
\be \label{eq:jexpand} J(z_c)\equiv J_0(z_c)+\varepsilon J_1(z_c)+\dots
\ee 
Hence
\be
X(z_{c_1}^-)=\left(1+2\pi i \varepsilon\, \text{Res}_{z_2}\,F\right)J(z_{c_1})\chi_{\rm out}(z_{c_1}^-)~.
\ee
Next, integrate around the point $z_1$ up to the second crossing point $z_{c_2}^-$, producing
\begin{align}
X(z_{c_2}^-)&=\left(1+2\pi i \varepsilon\, \text{Res}_{z_2}\,F\right)J(z_{c_1})\left(1+2\pi i \varepsilon\, \text{Res}_{z_1}\,\tilde{F}+\varepsilon\int_{z_{c_1}^-}^{z_{c_2}^-} \tilde{F}\right)\tilde{V}\nonumber\\
&=\left(1+2\pi i \varepsilon\, \text{Res}_{z_2}\,F\right)J(z_{c_1})\left(1+2\pi i \varepsilon\, \text{Res}_{z_1}\tilde{F}\right)\chi_{\rm out}(z_{c_2}^-)~.
\end{align} 
Finally, to get the monodromy matrix $M$, we match once more across $z_{c_2}$, resulting in
\be\label{eq:mon1}
M=\left(1+2\pi i \varepsilon\, \text{Res}_{z_2}\,F\right)J(z_{c})\left(1+2\pi i \varepsilon\, \text{Res}_{z_1}\tilde{F}\right)J^{-1}(z_c)~.
\ee
It is then easy to check by plugging (\ref{eq:jexpand}) into (\ref{eq:mon1}) (and using $J^{-1}=J_0^{-1}-\varepsilon J_0^{-1}J_1 J_0^{-1}+\dots$) that $J_1$ does not contribute to $M$ at $\mathcal{O}(\varepsilon)$. To leading order, the monodromy matrix is
\be
M=1+2\pi i\varepsilon\left(\text{Res}_{z_2}\,F+J_0(z_c)(\text{Res}_{z_1}\tilde{F})J_0^{-1}(z_c)\right)~.
\ee
Now to compute the identity block we must impose trivial monodromy, which means solving
\be\label{eq:deltamonauto}
\text{Res}_{z_2}\,F+J_0(z_c)\text{Res}_{z_1}\tilde{F}J_0^{-1}(z_c)=0
\ee
for the $b_i$.  The next step is to solve the differential equation
\begin{equation}\label{eq:diffeqblockexp}
\frac{\partial f_0}{\partial z_i}=\frac{6h_\mathcal{Q}}{c}b_i=c_i^{\mathcal{Q}}~\quad\quad i=1,2~.
\end{equation}
By symmetry, the dominant path $\gamma$ will cross the heavy insertions at $z_c=i(1-\sigma)$. The solution to (\ref{eq:diffeqblockexp}) in the $\sigma\rightarrow0$ limit is 
\be\label{fzzz}
f_{0}=\frac{12 h_\mathcal{Q}}{c}\log\left[\left\lbrace1-i\rho-i(1+i\rho)z_1\right\rbrace z_2^{\frac{1}{2}(1-i\rho)}-e^{\frac{\pi \rho}{2}}\left\lbrace1+i\rho -i(1-i\rho)z_1\right\rbrace z_2^{\frac{1}{2}(1+i\rho)}\right]+\text{const}.
\ee
The constant is fixed by demanding that $f_0$ give the correct behavior for a vacuum correlator as $z_2\rightarrow i$, that is $f_0\sim\frac{12 h_\mathcal{Q}}{c} \log (z_1-i)$.  This fixes
\be
\text{const}.=\frac{12 h_\mathcal{Q}}{c}\log\left(\frac{e^{-\frac{\pi}{4}(i+\rho)}}{2 \rho}\right) \ .
\ee
The holomorphic identity block is $\mathcal{F}_0=\exp\left(-\frac{c}{6}f_0\right)$. To go from this to (the dominant contribution to) the Euclidean correlator, we simply multiply it by the analogous anti-holomorphic contribution:
\begin{equation}
G(z_i,\bar{z}_i)\simeq{\cal F}_0(z_{i})\bar{{\cal F}}_0(\bar z_{i})\simeq\exp\left(-\frac{c}{6} f_0(z_1,z_2)-\frac{c}{6} \bar{f}_0(\bar{z}_1,\bar{z}_2)\right)~.
\end{equation}
This is the answer on the Euclidean plane. We are actually interested in the correlation function on the cylinder, which means we must invert the map $w\mapsto z=e^w$. This gives a Jacobian factor in $G(w_i,\bar{w}_i)$:
\be
G(w_i,\bar{w}_i)=e^{h_{\mathcal{Q}}(w_1+\bar{w}_1+w_2+\bar{w}_2)}\exp\left(-\frac{c}{6} f_0\left(e^{w_1},e^{w_2}\right)-\frac{c}{6} \bar{f}_0\left(e^{\bar{w}_1},e^{\bar{w}_2}\right)\right)\, .
\ee 
Now to obtain the Lorentzian correlator, we take $w_i=i\pi/2 +\tau_i$ and $\bar{w}_i=-i\pi/2+\tau_i$, then continue to Lorentzian times $\tau_i\rightarrow i t_i$. The final result is:
\be\label{circlefinal}
G(t_1,t_2)=i^{-2\Delta_{\cal Q}}\left(\frac{2}{\rho}\cos\left(\frac{t_1}{2}\right)\sinh\left(\frac{\rho\,t_2}{2}\right)-2\sin\left(\frac{t_1}{2}\right)\cosh\left(\frac{\rho\,t_2}{2}\right)\right)^{-2\Delta_\mathcal{Q}}~.
\ee
This is the autocorrelation function of an operator of dimension $\Delta_{\mathcal{Q}}$ in a CFT on a circle of radius $R=1$, with $t_1$ and $t_2$, respectively, before and after a global Vaidya quench. 

From this expression we can read off the answer for a CFT on an infinite line by reintroducing the radius of the circle $R$ and taking the limit $R\rightarrow \infty$. Before taking the limit, let us briefly discuss the interpretation of $\rho=\sqrt{4K-1}$.  In (\ref{gravityIdent}) we related $K$ to the mass of the final state black hole in the bulk dual, meaning that $\rho$ is the bulk dimensionless temperature $\rho=2\pi\ell/\beta_{\rm bulk}$. Via the usual AdS/CFT dictionary, we should then identify $\rho=2\pi R/\beta$ in the CFT, with $\beta$ the temperature of the late time equilibrium state. We can now take the $R\rightarrow\infty$ limit:
\begin{align}\label{linefinal}
G_{\rm line}(t_1,t_2)&= i^{-2\Delta_{\cal Q}} \lim_{R\rightarrow\infty}\left(\frac{2R}{\rho}\cos\left(\frac{t_1}{2R}\right)\sinh\left(\frac{\rho\,t_2}{2R}\right)-2R\sin\left(\frac{t_1}{2R}\right)\cosh\left(\frac{\rho\,t_2}{2R}\right)\right)^{-2\Delta_\mathcal{Q}}~,\nonumber\\
&=i^{-2\Delta_{\cal Q}}\left(\frac{\beta}{\pi}\sinh\left(\frac{\pi\,t_2}{\beta}\right)-t_1\cosh\left(\frac{\pi\,t_2}{\beta}\right)\right)^{-2\Delta_\mathcal{Q}}\, .
\end{align}
This matches precisely with the geodesic calculation on the gravity side in \cite{Balasubramanian:2012tu}.\footnote{The prefactor $i^{-2\Delta_{\cal Q}}$ does not appear in \cite{Balasubramanian:2012tu}. This is due to a different choice of operator normalization. We have normalized operators so $\langle O(z_1)O(z_2) \rangle = |z_1-z_2|^{-2\Delta}$ on the plane, whereas operators in  \cite{Balasubramanian:2012tu} are normalized so that $\langle O(w_1) O(w_2) \rangle \sim |w_1 - w_2|^{-2\Delta}$ as $w_1 \to w_2$ on the cylinder.}

In the analytic continuation to Lorentzian signature, we implicitly chose a prescription for analytically continuing past branch cuts. (In the finite-$n$ correlator, these appear whenever the probes hit the lightcones of the dust operators.) This choice of analytic continuation is equivalent to a choice of ordering for timelike separated operators (see \cite{Hartman:2015lfa} for a detailed review).
 The prescription we chose above, \textit{i.e.}, the na\"ive analytic continuation of \eqref{fzzz} without inserting any additional factors of $z_2 \to e^{2\pi i} z_2$, corresponds to the operators ordered as written in \eqref{eq.expectationValueVaidya}.  This out-of-time-order correlator has the appropriate ordering for expectation values in the state $|\V\rangle$.

\subsection{Entanglement Entropy}
We now move on to our second example: the entanglement entropy growth of an interval of length $L$ in the Vaidya state. The ingredients of the calculation are very similar to those in the last section, so we will be more brief. The result can also be interpreted as an equal-time spatial correlation function $G(x_1, t; x_2, t)$ of probe operators.

Our goal is to compute the entanglement entropy of an interval of length $L$ in the the state $|\mathcal{V}\rangle$. To do so, we use the usual replica trick and compute a correlation function of twist operators $G_\alpha(z_1,z_2)=\langle\mathcal{V}|\sigma_\alpha(z_1)\tilde{\sigma}_\alpha(z_2)|\mathcal{V}\rangle$ where $\sigma_\alpha$ and $\tilde{\sigma}_\alpha$ are conformal primaries of dimension 
\be
h_\alpha=\frac{c}{24}\left(\alpha-\frac{1}{\alpha}\right)~.
\ee
The entanglement entropy is related to the correlation function of twist operators via
\be
S_{EE}=\lim_{\alpha\rightarrow1}\frac{1}{1-\alpha}\log G_\alpha(z_1,z_2)\, .
\ee
\begin{figure}
\begin{center}
\includegraphics[width=0.6\textwidth]{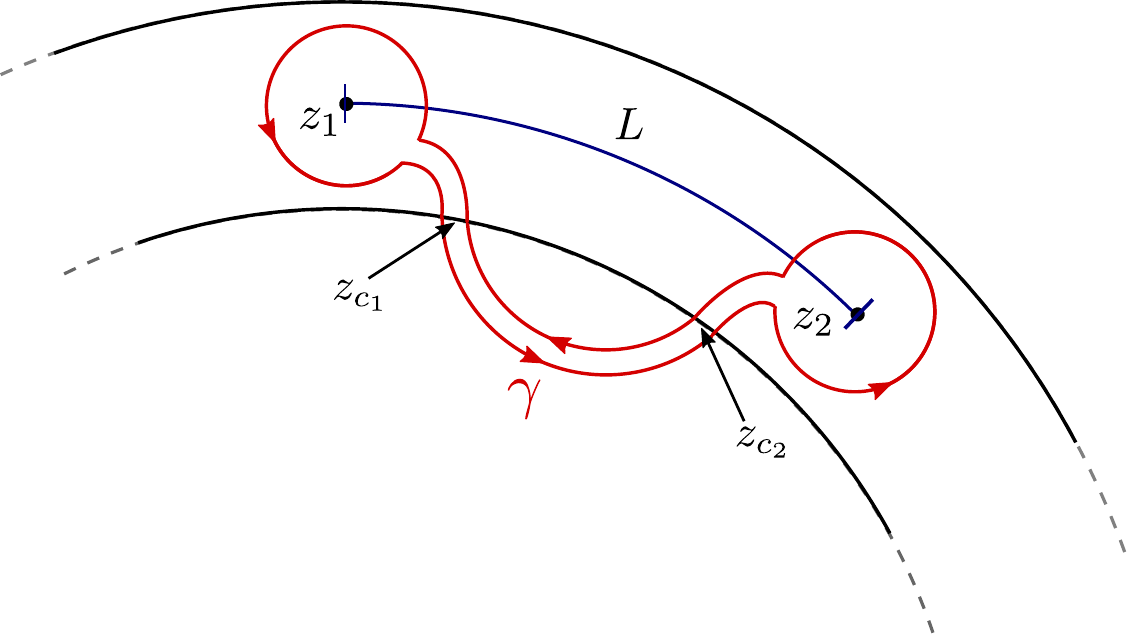}
\caption{Path on the $z$ plane which defines the Euclidean OPE channel in the calculation of entanglement entropy. The solid lines are the boundaries of the annulus $1-\sigma < |z| < 1 + \sigma$. We impose trivial monodromy on the path $\gamma$ to calculate the block in a given channel. The dominant contribution is obtained by maximizing over $z_{c_1}$, $z_{c_2}$. \label{fig:mononstromy}}
\end{center}
\end{figure}
We need to compute the monodromy matrix $M$ of a solution to (\ref{eq.monodromicODE}) around the path $\gamma$ shown in figure \ref{fig:mononstromy}. It is not difficult to see that the monodromy matrix for this path is 
\be
M=\left(1+2\pi i \varepsilon\, \text{Res}_{z_1}\,F\right)J(z_{c_1})AJ^{-1}(z_{c_2})\left(1+2\pi i \varepsilon\, \text{Res}_{z_2}F\right)J(z_{c_2})A^{-1}J^{-1}(z_{c_1})~
\ee
where $A$ is the matrix that integrates the solution $\chi_{\rm out}$ from $z_{c_1}$ to $z_{c_2}$. However, $A$ can be written as $A= 1+ \varepsilon \delta A$, and it is straightforward to show that $\delta A$ drops out of the expression for $M$ at $\mathcal{O}(\varepsilon)$. Therefore to leading order
\be
M=1+2\pi i \varepsilon \left[\text{Res}_{z_1}\,F+J_0(z_{c_1})J_0^{-1}(z_{c_2})(\text{Res}_{z_2}\,F)\,J_0(z_{c_2})J_0^{-1}(z_{c_1})\right]~.
\ee
Until now we have treated the crossing points $z_{c_1}$ and $z_{c_2}$ as arbitrary. However, in computing the dominant contribution to the correlator, we are instructed to maximize the final answer over these crossing points. One can argue by symmetry that the dominant path $\gamma$ should be symmetric about its middle, that is for $z_{c_1}=(1-\sigma)e^{i q}$ then $z_{c_2}=(1-\sigma)e^{i(L-q)}$ with $q\in [0,L/2]$. A similar phenomenon happens for geodesics in Vaidya---those that cross the shell of null dust are symmetric about the middle. It is satisfying to find a similar condition arise in CFT.

Following the procedure outlined previously, we solve for the $b_i$ that set $M=1_{2\times 2}$ and thereafter integrate them to obtain $f_0$. We find (labeling $z_i=e^{i\theta_i}$):
\begin{multline}
f_0\left(e^{i\theta_1},e^{i\theta_2}\right)=\frac{6 h_\alpha}{c}\log\Bigg[-4e^{(L+2q)(i+\rho)+i(\theta_1+\theta_2)}\bigg\lbrace2(\rho+1)^2\sin\left(\frac{L}{2}-q\right)\cosh\left[\frac{\rho}{2}(L-\theta_1-\theta_2)\right]\\
-i(\rho-i)^2\sinh\left[\left(\frac{L}{2}-q\right)(\rho+i)+\frac{\rho}{2}(\theta_1-\theta_2)\right]\\
+i(\rho+i)^2\sinh\left[\left(\frac{L}{2}-q\right)(\rho-i)+\frac{\rho}{2}(\theta_1-\theta_2)\right]\bigg\rbrace^2\Bigg]+\text{const}.
\end{multline}
We again fix the integration constant by demanding that the block give the vacuum answer when $\gamma$ lies entirely outside of the strip, that is $f_0=\frac{12h_\alpha}{c}\log\left[\frac{\sin\left(\frac{L}{2}-q\right)}{\epsilon_{UV}/2}\right]$ for $\theta_1=q$ and $\theta_2=L-q$. Here $\epsilon_{UV}$ is a UV cutoff that regulates the definition of the twist operator \cite{Calabrese:2004eu}.

Finally, the entanglement entropy is 
\be
S_{EE}=\lim_{\alpha\rightarrow 1}\frac{1}{1-\alpha}\left(-\frac{c}{6}f_0\left(e^{w_1},e^{w_2}\right)-\frac{c}{6}\bar{f}_0\left(e^{\bar{w}_1},e^{\bar{w}_2}\right)+h_\alpha(w_1+\bar{w}_1+w_2+\bar{w}_2)\right)~.
\ee
with $w_1=i \theta_1$ and $w_2=i(\theta_1+L)$. To continue to Lorentzian times we simply take $\theta_1=t$. We are not yet done, as we still need to maximize $S_{EE}$ over the free parameter $q$ labeling the point where $\gamma$ crosses through background insertions. This cannot be solved in closed form for $q$, however we can solve $\partial S_{EE}/\partial q=0$ for $t$ and obtain a parametric expression for the entanglement entropy growth of the interval. Once the dust settles we find (for $q\in[0,L/2]$):
\begin{align}\label{eq.FullEE}
t&=\frac{\beta}{2\pi}\cosh^{-1}\left\lbrace\cosh\left(\frac{2\pi q}{\beta}\right)+\frac{2\pi R}{\beta} \tan\left(\frac{\frac{L}{2}-q}{R}\right)\sinh\left(\frac{2\pi q}{\beta}\right)\right\rbrace~,\\
\!\!\!\!\!\!S_{EE}&=\frac{c}{3}\log\left\lbrace\frac{R\sin\left(\frac{\frac{L}{2}-q}{R}\right)\cosh\left(\frac{2\pi q}{\beta}\right)+\frac{\beta}{2\pi}\left[1+\frac{1}{2}\left\lbrace 1+\left(\frac{2\pi R}{\beta}\right)^2\right\rbrace\tan^2\left(\frac{\frac{L}{2}-q}{R}\right)\right]\cos\left(\frac{\frac{L}{2}-q}{R}\right)\sinh\left(\frac{2\pi q}{\beta}\right)}{\epsilon_{UV}/2}\right\rbrace~.\notag
\end{align}
We have reintroduced the radius of the CFT circle  $R$ in the final answer and replaced  $\rho$ with $2\pi R/\beta$. This answer was calculated via a bulk geodesic length in  \cite{Hubeny:2007xt,Ziogas:2015aja}. Taking the $R\rightarrow \infty$ limit of the above answer gives the planar Vaidya geodesic length calculated in \cite{Balasubramanian:2011ur,Balasubramanian:2010ce}.

This formula for the growth of entanglement after the Vaidya quench is only valid for $0<L<\pi R$. For $L>\pi R$ one simply replaces $L\rightarrow 2\pi R-L$ in the above formula, implying that the entanglement entropy of the interval of length $L$ is equal to the entanglement entropy of its complement, as expected in a pure state.

\section{Discussion of Information Loss}

The exact CFT calculation is obviously unitary, but the leading term in the $1/c$ expansion at early times may not be.  In fact, since it agrees with the gravity side, we expect the telltale signs of information loss in the approximate CFT calculations. In eigenstates,
this was demonstrated for 2-point functions in \cite{Fitzpatrick:2014vua} (see also \cite{Fitzpatrick:2015zha,Fitzpatrick:2015foa}), and discussed in terms of entanglement entropy in \cite{Asplund:2014coa}.  The story for black holes forming by dynamical collapse is similar. Information is lost at large $c$, but restored by non-perturbative corrections in the $1/c$ expansion. Such a picture for information loss and recovery is expected from general arguments ---  it has been observed in toy models for the information paradox, such as matrix quantum mechanics \cite{Iizuka:2008hg}, and related behavior can be argued to occur in large-$N$ gauge theory \cite{Festuccia:2006sa}. Here we confirm this expectation for our detailed model of the 3d black hole.

\bigskip \noindent \textbf{Correlators} \ \\
 Consider the late-time behavior of the correlator \eqref{circlefinal}:
\be\label{gexp}
G(t) \sim \exp\left(-\frac{2\pi \Delta_{\mathcal{Q}} t}{\beta}\right)  \ ,
\ee
where we have set $t_1=0, t_2=t$. This permanent exponential decay is incompatible with quantum mechanics, as pointed out in the case of the eternal black hole by Maldacena \cite{Maldacena:2001kr}. This follows on general grounds for any system with finite entropy (see for example \cite{Harlow:2014yka}).  Intuitively the reason is that in a pure state $|\Psi\rangle = \sum_n a_n |n\rangle $, if we decompose the correlator as a sum over eigenstates,
\be\label{eigend}
G(t) \equiv \langle \Psi | O(t) O(0) |\Psi\rangle = \sum_{n,k} e^{i (E_n-E_k) t} a_n^* \langle n | O | k \rangle \langle k | O |\Psi\rangle \ ,
\ee
then the large phases in this sum at late times can make the correlator very small, but cannot cancel exactly.

Returning to the first step in the CFT calculation, it is obvious what went wrong -- we kept only a single term in the conformal block expansion  \eqref{exactOPE}. Under our assumptions about the spectrum, this term is exponentially dominant at early times, but it cannot be the full answer, since the vacuum block alone (or indeed any individual conformal block) violates crossing symmetry. In general, it is not possible to compute the subleading terms, which come from heavy operator exchange and depend on the details of the CFT. But we can easily see from crossing symmetry that they must exist, and dominate at late times. The decomposition \eqref{eigend} in the state $|\Psi\rangle = |\V\rangle$ can be viewed as an OPE channel,
\begin{equation}\label{dualOPE}
G = 
\sum_{\mbox{\footnotesize primaries}}\quad
\begin{gathered}
\includegraphics[width=300px]{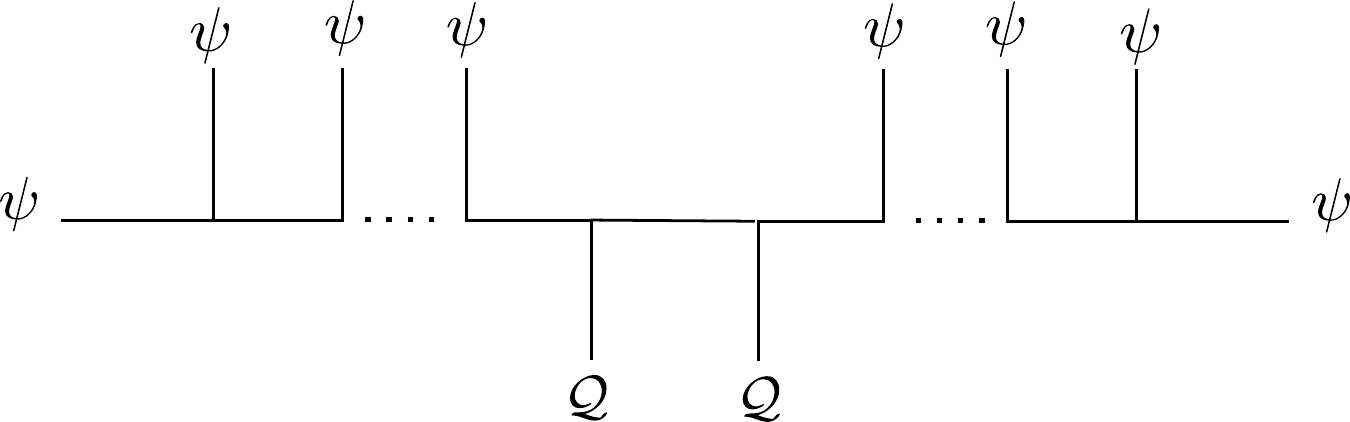}
\end{gathered} \ ,
\end{equation}
where the $\psi$'s on the left half of the diagram are the ones inserted at $|z|=1+\sigma$, and the $\psi$'s on the right are those inserted at $|z|=1-\sigma$. Then the same general reasoning that applies to \eqref{eigend} also applies to this correlator, so it cannot go exponentially to zero. Crossing symmetry then implies that the heavy operators in the original channel \eqref{exactOPE} produce a finite, late-time tail that resolves the tension with unitarity. At early times, the contributions of these heavy operators are suppressed nonperturbatively in $1/c$.

Of course, this does not explain how information is recovered in the bulk --- Hawking's paradox is a problem with bulk effective field theory, so must ultimately be solved on the gravity side. It does, however, sharpen the problem, since in the CFT (unlike in Hawking's calculation) we made a controlled approximation to a well defined exact calculation, and confirmed that this approximation breaks down before unitarity is violated. In gravity language, this supports the standard expectation that information should be restored by effects nonperturbative in $G_{\rm N}$.

\bigskip \noindent \textbf{Entanglement entropy} \ \\
Entanglement entropy is also a delicate probe of unitarity.  In a pure state, 
\be\label{sac}
S_A = S_{A^C}
\ee
where $A^C$  is the complement of region $A$.  Even Hawking's original calculation of black hole evaporation in asymptotically flat spacetime can be viewed as a violation of \eqref{sac}, taking region $A$ to be a portion of null infinity. In this case $A$ contains the early Hawking radiation, and $A^C$ contains the late Hawking radiation, so \eqref{sac} holds if the total state of the radiation is pure.

In our case, region $A$ is a segment of the CFT circle at fixed time.  The calculations of $S_A$ and $S_{A^C}$ are obviously identical, since they both correspond to the same twist correlator.  On the bulk side, this means that the answer we have derived allows the entanglement geodesic to be taken through the black hole horizon.  For an eternal black hole, this would be disallowed by the homology condition for the extremal surface, conjectured in \cite{Ryu:2006bv} and derived in \cite{Lewkowycz:2013nqa}.  However, for a collapsing black hole, the homology condition (in this case only a conjecture, since \cite{Lewkowycz:2013nqa} does not apply) allows us to deform the extremal surface into the past, behind the formation of the horizon, and onto the other side of the black hole \cite{Takayanagi:2010wp}.  The choice of channels in the CFT calculation directly mimics this procedure and confirms this expectation directly from CFT. The bulk geodesic that goes the `long way' around the horizon corresponds to the identity block in a subdominant OPE channel of the CFT; these two channels exchange dominance when region $A$ is exactly half the system size, $L = \pi R$.

\bigskip

\bigskip

\textbf{Acknowledgments} \ \ We thank Arjun Bagchi, Alice Bernamonti, Alejandra Castro, Federico Galli, John Cardy, Steven Gubser, Nima Lashkari, Hong Liu, Eric Perlmutter, Mukund Rangamani, Dan Roberts, Vyacheslav Rychkov, Edgar Shaghoulian, Douglas Stanford,  and Sasha Zhiboedov for useful comments and conversations. TA is supported in part by NSF grant PHY-0967299 and by the U.S. Department of Energy under grant Contract Number DE-SC0012567. TH is supported by DOE grant DE-SC0014123. JS and AR are supported by the Fonds National Suisse de la Recherche Scientifique (FNS) under grant number 200021\_162796 and by the NCCR 51NF40-141869 ``The Mathematics of Physics'' (SwissMAP). The work of AR is also supported in part by the Belgian American Educational Foundation. TH and JS thank the KITP for support during the programs \textit{Entanglement in Strongly-Correlated Quantum Matter} and \textit{Quantum Gravity Foundations: UV to IR}, funded by the National Science Foundation under Grant No.~NSF PHY11-25915. 

\appendix
\section{Sources vs. States}\label{ap:sourcevstate}

\begin{figure}
\begin{center}
\includegraphics[width=0.5\textwidth]{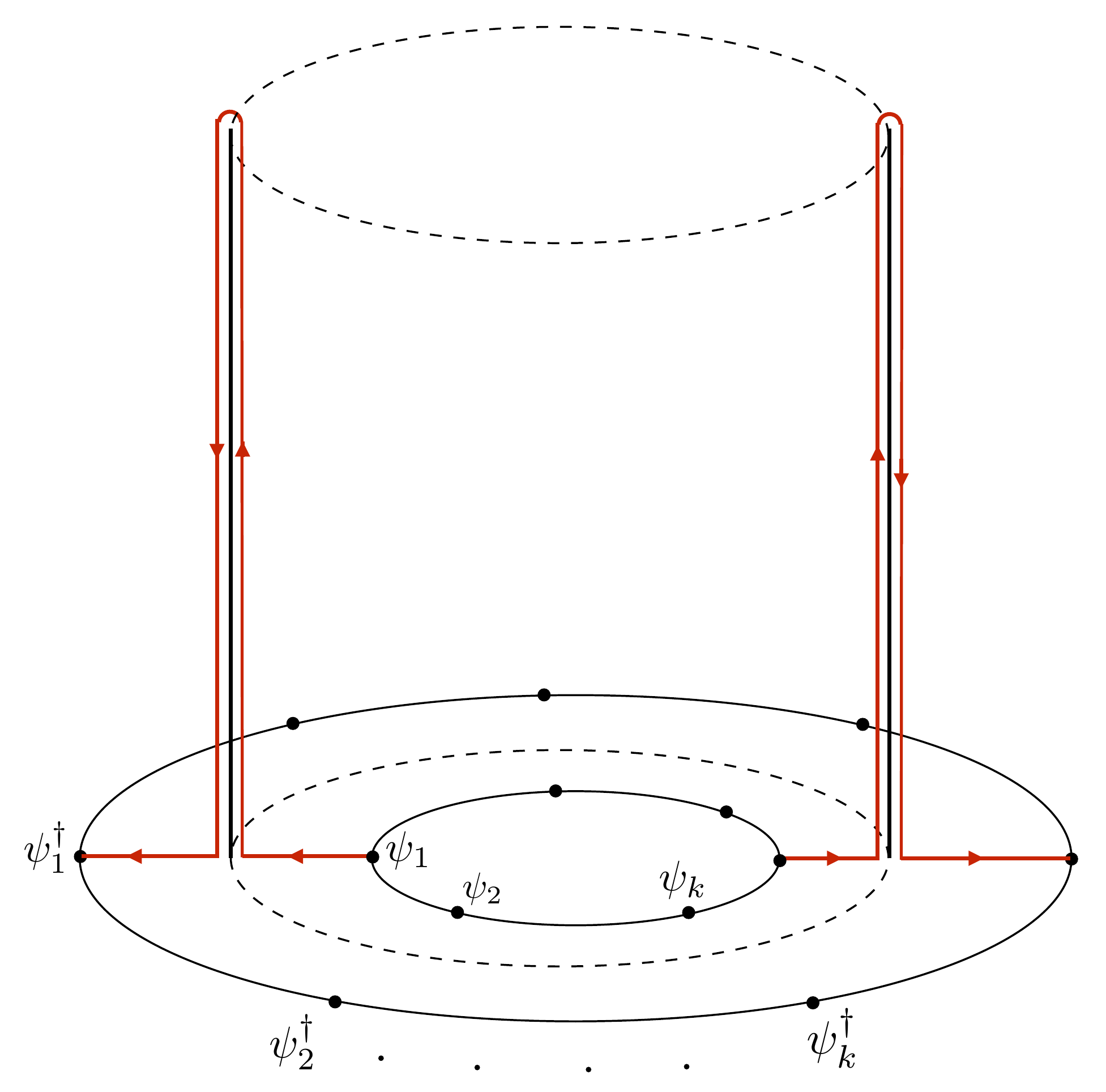}
\caption{\small The CFT calculation dual the null shell collapse in AdS$_3$ gravity employs the in-in formalism. The time evolution contour is indicated in red: The state is prepared by a Euclidean path integral on the unit disk, with operators inserted at $|z|=1-\sigma$, followed by forward-backward evolution along the Lorentzian part of the contour. The final part of the evolution is over the outside of the Euclidean disk, with operators inserted at $|z|=1+\sigma$. In practice we construct all quantities via analytic continuation from the Euclidean block. 
\label{fig:CFTstate}}
\end{center}
\end{figure}

\renewcommand{\O}{\mathcal{O}}
 The reader may wonder whether it is more natural to perform a different calculation: Instead of inserting primary operators offset in Euclidean time, we could instead add to the CFT a source term deforming the action,
 \be\label{eq.operatorDeformation}
 S \rightarrow S + \int d^2x \, J(t,x)\O(t,x)\,,
 \ee
where $\O$ is a scalar operator and $J(t,x)$ is its classical source. Performing a spatially homogeneous quench means that $J = J(t)$, and, following the gravity calculation of \cite{Bhattacharyya:2009uu}, we take $J(t)$ to be compactly supported in time, that is we choose a smooth function $J(t)$ such that $J(t) = j_0 \neq 0 $ on an interval of size $\eta$ centered on $t = 0$ and zero otherwise. After the source turns off, the system is simply the original CFT in some excited state.  

Under our assumption that correlators are dominated by stress tensor exchange, all that matters is the value of $\langle T\rangle$ in this excited state -- if it agrees with $\langle \V| T | \V\rangle$ computed in section \ref{s:cfttech}, then all probe observables will agree in these two approaches.
The calculation with sources appears to be more difficult, however, since a finite, exponentiated source produces UV divergences that need to be regulated and resummed.  We will not attempt the full calculation, but in what follows, we describe the setup in the approach \eqref{eq.operatorDeformation} and check that the leading term for $|j_0| \ll 1$ --- the collapse of a small mass black hole, for which resummation is not necessary --- agrees with our calculations in the state $|\V\rangle$. 

In the presence of a source \eqref{eq.operatorDeformation}, correlation functions are computed in the interaction picture as
\be\label{eq.inincorrelator}
\langle {\cal Q}_1(t_1)\cdots {\cal Q}_p(t_p)\rangle = \langle  U^\dagger(t,-\infty) {\cal Q}^I_1(t_1)\ldots {\cal Q}_p^I(t_p) U(t,\infty) \rangle\,,
\ee
where $t = {\rm max}(t_1,\ldots t_p)$ is the largest time of any of the operator insertions and the superscript `$I$' denotes that the corresponding operator is in the interaction picture with respect to the decomposition (\ref{eq.operatorDeformation}). The evolution operator is
\be
U(t_B, t_A)  = T\exp\left(-i \int_{t_A}^{t_B}H_{I}(t')dt'\right) \ .
\ee
Such amplitudes are computed in the `in-in' formalism, starting and ending in the CFT vacuum. This follows from the perturbative expansion of (\ref{eq.inincorrelator}). In essence the time evolution operators in  (\ref{eq.inincorrelator}) prescribe a sum over different time orderings of the operators, each of which can be reconstructed using a suitable $\epsilon$ prescription from the Euclidean correlation function. One may similarly view the `state' computation as an in-in correlator, whereby the state is produced by Euclidean evolution for a time $\sigma$ (the insertions on the circle of radius $1-\sigma$) before switching to Lorentzian evolution to compute the expectation value for ${\cal Q}(t)$. The overlap with the conjugate state $\langle {\cal V}|$ then corresponds to backwards time evolution, as for the `in-in' prescription  (see Fig. \ref{fig:CFTstate}).

Let us illustrate the procedure following from the expression (\ref{eq.inincorrelator}) by computing the expectation value of the stress tensor to leading order in the perturbative expansion. We focus on the second-order contribution
\be\label{eq.Q2ndOrder}
\langle {\cal Q}(t) \rangle_{(2)} = -\int_{-\infty}^tdt_2 \int_{-\infty}^{t_2}dt_1 \langle   \left[ H_I (t_1), \left[ H_I(t_2),{\cal Q}(t) \right] \right]  \rangle J(t_1)J(t_2) \,,
\ee
where the quench Hamiltonian is
\be
H_I(t) = \int dy J(y,t) \O(y,t)
\ee
and ${\cal  Q}(t)$ is the operator whose time evolution we wish to determine. The term (\ref{eq.Q2ndOrder}) will be the leading contribution when zeroth and first-order contributions vanish.
%
If we are interested in energy density, we should take ${\cal Q} = T_{00}$. Then this integral contains a U.V. divergence $\sim (2\sigma)^{2-2\Delta}$, where $\sigma$ is a regulator (see \cite{Faulkner:2014jva}), and $\Delta$ is the conformal dimension of $\O$. It follows that $\langle {\cal Q}\rangle \sim \Theta(t) \eta^{2-2\Delta} + {\rm U.V.}$, which coincides with the result for a marginal operator in the gravity calculation of \cite{Bhattacharyya:2009uu}.

Similarly we can compute the entanglement entropy of an interval of size $L$ with endpoints $\ell_1$ and $\ell_2$, in which case we take ${\cal Q}(t) = \sigma_n(t,\ell_1)\tilde \sigma_n(t,\ell_2)$ in (\ref{eq.Q2ndOrder}). Now the leading contribution to the R\'enyi entropy near $n=1$ comes from the heavy-heavy-light-light four-point function $\langle \O \sigma_n \tilde\sigma_n \O \rangle$, which was computed in \cite{Asplund:2014coa}, suitably continued to reproduce the Lorentzian orderings in (\ref{eq.Q2ndOrder}). From this one can recover the entanglement entropy in the limit $n\rightarrow 1$. We have calculated the resulting double integral over sources numerically and found agreement with the full answer (\ref{eq.FullEE}) to leading order in the small-mass expansion $EL \ll 1$, where $E \propto j_0^2/\eta^{2\Delta -2}$ is the energy of the final black hole. Such energy scaling has previously been pointed out by \cite{Das:2014jna,Berenstein:2014cia}.

\end{spacing}

\bibliographystyle{utphys}
\bibliography{3DBlackHoleRefs}{}

\end{document}